\documentclass[aps,prd,groupedaddress,floatfix,longbibliography,superscriptaddress]{revtex4-1}
\usepackage{amsmath}
\usepackage{amssymb}
\usepackage{amsfonts}
\usepackage{graphicx}
\usepackage{bbold}
\usepackage{bm}
\usepackage{bm}
\usepackage{times,float}
\usepackage{graphicx}
\usepackage[usenames,dvipsnames,svgnames]{xcolor}
\usepackage{hyperref}
\hypersetup{colorlinks=true, linkcolor=Blue, citecolor=Magenta,urlcolor=Blue}

\begin{document}

\title{ Scalar Casimir effect for a conducting cylinder in a Lorentz violating background }

\author{A. M. Escobar-Ruiz}
\email{admau@xanum.uam.mx}
\affiliation{Departamento  de  F\'isica,  Universidad  Aut\'onoma  Metropolitana-Iztapalapa, San Rafael Atlixco 186, 09340 Ciudad de M\'{e}xico, M\'{e}xico}

\author{A. Mart\'{i}n-Ruiz}
\email{alberto.martin@nucleares.unam.mx}
\affiliation{Instituto de Ciencias Nucleares, Universidad Nacional Aut\'{o}noma de M\'{e}xico, 04510 Ciudad de M\'{e}xico, M\'{e}xico}

\author{C. A. Escobar}
\email{carlos\_escobar@fisica.unam.mx}
\affiliation{Departamento  de  F\'isica,  Universidad  Aut\'onoma  Metropolitana-Iztapalapa, San Rafael Atlixco 186, 09340 Ciudad de M\'{e}xico, M\'{e}xico}
%\affiliation{Instituto de F\'{i}sica, Universidad Nacional Aut\'{o}noma de M\'{e}xico, Apartado Postal 20-364, Ciudad de M\'{e}xico 01000, M\'{e}xico}

\author{Rom\'an Linares}
\email{lirr@xanum.uam.mx}
\affiliation{Departamento  de  F\'isica,  Universidad  Aut\'onoma  Metropolitana-Iztapalapa, San Rafael Atlixco 186, 09340 Ciudad de M\'{e}xico, M\'{e}xico}

\begin{abstract}
Following a field-theoretical approach, we study the scalar Casimir effect upon a perfectly conducting cylindrical shell in the presence of spontaneous Lorentz symmetry breaking. The scalar field is modeled by a Lorentz-breaking extension of the theory for a real scalar quantum field in the bulk regions. The corresponding Green's functions satisfying Dirichlet boundary conditions on the cylindrical shell are derived explicitly. We express the Casimir pressure (i.e. the vacuum expectation value of the normal-normal component of the stress-energy tensor)
%Since the two-point Green's function is defined as the vacuum expectation value of the time-ordered product of two fields, the Casimir pressure (i.e. the vacuum %expectation value of the normal-normal component of the stress-energy tensor) can be written
% 
as a suitable second-order differential operator acting on the corresponding Green's functions at coincident arguments. The divergences are regulated by making use of zeta function techniques, and our results are successfully compared with the Lorentz invariant case. Numerical calculations are carried out for the Casimir pressure as a function of the Lorentz-violating coefficient, and an approximate analytical expression for the force is presented as well. It turns out that the Casimir pressure strongly depends on the Lorentz-violating coefficient and it tends to diminish the force.
\end{abstract}

\maketitle

\section{Introduction}

Our current description of nature is based on the standard model of particle physics and general relativity, although they operate at very different energy scales. Both theories have been successful in explaining and predicting the observed physical phenomena with a high degree of accuracy, however they come with their own set of deficiencies: ultraviolet divergences in quantum field theories and singularities in general relativity. It is widely believed that a unified quantum theory of gravity will solve these problems, and this justifies the great effort of theoretical physicists in this direction in the last decades. To date, only candidates have been proposed, such as non-commutative spacetime theories \cite{Pospelov, Carroll, Carlson, Anisimov}, string theories \cite{Samuel, Potting, Hashimoto} and  loop quantum gravity \cite{Klinkhamer, Ashtekar, Rovelli, Ashtekar2, Jacobson}. Within these proposals we find some mechanisms that lead to a breakdown of Lorentz symmetry, which is expected to occur at distances of the order of the Planck length $\ell  _{p}$, where the quantum nature of the spacetime becomes important. Evidently, a better understanding on the consequences of the breakdown of Lorentz invariance at scales larger than $\ell  _{p}$ would provide valuable information about the microscopic structure of spacetime. Nowadays, investigations concerning Lorentz violation are mostly conducted under the framework of the Standard-Model Extension (SME) \cite{Kostelecky1, Kostelecky2}, which is an effective field theory that contains the standard model, general relativity, and all possible operators that break Lorentz invariance. 

The Casimir effect \cite{Casimir} is one of the most remarkable consequences of the nonzero vacuum energy predicted by quantum field theory. Generally speaking, it refers to the stress on bounding surfaces when a quantum field is confined to a finite volume of space. The origin of the stress is the suppression of some quantum field modes due to the boundary conditions on the surfaces. The boundaries can be material media, interfaces between two phases of the vacuum, or topologies of space. In any case, the modes of the quantum fields are restricted, giving rise to a macroscopically measurable force \cite{Bressi}. The fundamental character of the Casimir effect, plus its current experimental accuracy \cite{Lamoreaux, Mohideen, Roy, Mostepanenko}, offers a great opportunity to test any theory, specially whose that introduce new and/or non standard ideas. Also, the Casimir effect stands as a potential handle to distinguish between Lorentz-invariant and Lorentz-violating formulations of quantum field theory. In this regard, the paradigmatic Casimir effect between two parallel conductive plates has been extensively studied in different scenarios, including spacetimes with nontrivial topologies \cite{DEWITT,Ford}, non-Euclidean spacetimes \cite{Dowker1978, Aliev, HUANG}, string theory \cite{FABINGER, Holger, Damour}, and theories with minimal length \cite{HARBACH, Nouicer2005, Polymer}, just to name a few. Concerning Lorentz-violating field theories, the Casimir effect has been attracted great attention as well. For example, the parallel plate configuration has been considered within the electromagnetic \cite{Kharlanov, Martin1, Martin2}, scalar \cite{Petrov1, Petrov2, Petrov3, Escobar1, Escobar2} and gravitational \cite{Santos2019} sectors of the minimal SME. The Casimir effect in the cylindrical and spherical geometries has also attracted great attention due to its simplicity and important applications \cite{MILTON198049, Fishbane, Graham, Graham2}. Recently, the scalar Casimir stress upon a spherical shell in $D+1$-dimensions has been investigated in Ref. \cite{Martin3}. However, the analysis of the Casimir effect for a cylindrical configuration in a Lorentz-violating field theory is still lacking. Our main goal here is to fill in this gap.

In this paper we study the scalar Casimir effect for a conducting cylinder in a Lorentz-violating (LV) background. The model is defined by the standard Klein-Gordon Lagrangian supplemented by the LV term $\lambda( u ^{\mu} \partial _{\mu } \phi ) ^{2}$, where $\lambda$ and the constant background four-vector $u ^{\mu} = ( u ^{0}, \vec{u} \,  )$ control the intensity and direction of the breakdown of Lorentz symmetry, respectively. We evaluate the pressure on the cylindrical shell by calculating the vacuum expectation value (VEV) of the normal-normal component (to the cylindrical boundaries) of the stress-energy tensor \cite{Miltonbook}.  By means of the point-splitting technique, we express the VEV as a second-order differential operator acting upon the corresponding bulk Green's functions \cite{Brown, Deutsch}. As usual, divergences arise, and we deal with by making use of the zeta function regularization technique \cite{Elizalde1, Elizalde2}. Finally, we evaluate numerically the expression for the Casimir pressure, and we also obtain an asymptotic  analytical expression which perfectly agrees with the numerical results. In short, the Lorentz-breaking term tends to diminish the pressure. 

The layout of the present work is as follows. In Sec. \ref{LV Sec} we introduce the Lorentz-violating model we deal with and some of its main properties. Section \ref{GFsection} is devoted to the derivation of the Green's functions that satisfy Dirichlet boundary conditions on the cylindrical shell. We consider two particular cases: the timelike case, defined by $u ^{0} = 1$ and $\vec{u} = \vec{0}$, and the radial spacelike case, defined by $u ^{0} = 0$ and $\vec{u} = \vec{e} _{\rho}$, being $\vec{e} _{\rho}$ the radial unit vector in the cylindrical coordinate system. In Sec. \ref{CEsection} we compute the vacuum stress-energy tensor by means of standard point-splitting  technique and obtain analytical expressions for the Casimir stress. Thereafter, in Sec. \ref{NR1} we resolve the divergence by means of the zeta function regularization technique, and we evaluate numerically the Casimir force. An asymptotic analysis of the expression allows to obtain an approximate expressions for the pressure, which correctly agrees with the numerical result. The summary of our results, concluding remarks and outlook is presented in Sec. \ref{Conclusection}.  The metric signature will be taken as $(+,-,-,-)$. We perform the derivation of the Casimir force using natural units $\hbar = c = 1$. The SI units will be restored at the end to present the final results.

\section{The model} \label{LV Sec}

The model we shall consider consists of the Klein-Gordon Lagrange density supplemented with a Lorentz-violating term of the form $\lambda \left( u ^{\mu} \partial _{\mu} \phi \right) ^{2}$, which was originally conceived within the scalar sector of the SME \cite{Kostelecky1, Kostelecky2}. The full Lagrange density for a LV massive real scalar field is \cite{Petrov1, Gomes}
\begin{align}
\mathcal{L}  = \frac{1}{2} \left[ \partial _{\mu} \phi \, \partial ^{\mu} \phi +  \lambda \left( u ^{\mu}  \partial _{\mu} \phi \right) ^{2} - m ^{2} \, \phi ^{2}  \right] ,  \label{Lagrangian1}
\end{align}
where $u ^{\mu} \equiv (u _{0} , \vec{u} \, )$ denotes a nonzero $(3+1)$-dimensional constant vector and the parameter $\lambda$ is such that $ \vert \lambda \vert < 1$, since the converse could introduce instabilities in the theory. The presence of the constant vector $u ^{\mu}$ in the second term of Eq. (\ref{Lagrangian1}) explicitly breaks Lorentz invariance, since it specifies privileged spacetime directions. Clearly, the Lorentz symmetric case is recovered as $\lambda=0$. Within the framework of effective field theories describing physics beyond the standard model, such as the SME, we expect the Lorentz-violating term to be very small, i.e. $\|  \lambda u^{\mu} u ^{\nu} \| \ll 1$. However, this kind of theories has been used to describe condensed matter systems, in which Lorentz symmetry is naturally broken, and hence the parameter $\lambda$ is not necessarily small.

Variation of the action $S= \int d ^{4} x \, \mathcal{L} $ produces a modified Klein-Gordon equation, which in terms of the $(3+1)$-dimensional D'Alembert operator $\Box = \partial _{\mu} \partial ^{\mu}$, reads
\begin{align}
\left[ \, \Box + \lambda  \left( u _{\mu}  \partial ^{\mu} \right) ^{2}  +  m ^{2} \right] \phi (x) = 0 . \label{EQLV}
\end{align}
In this paper we will be concerned with solutions to Eq. (\ref{EQLV}) satisfying Dirichlet boundary conditions (BCs) on a cylindrical shell of radius $R$ to evaluate the Casimir pressure, which is defined in terms of the stress-energy tensor. In the presence of the Lorentz-violating term, the corresponding stress-energy tensor is
\begin{align}
T ^{\mu \nu} = (\partial ^{\mu} \phi ) (\partial ^{\nu} \phi ) + \lambda  u ^{\mu} (\partial ^{\nu} \phi ) (u _{\sigma} \partial ^{\sigma} \phi ) - \eta ^{\mu \nu} \mathcal{L} , \label{Tmunu}
\end{align}
where $\eta ^{\mu \nu} = \textrm{diag} (1,-1,-1,-1)$ is the usual Minkowski flat spacetime metric in $(3+1)$-dimensions. As can be directly verified, the stress-energy tensor (\ref{Tmunu}) is conserved ($\partial _{\mu} T ^{\mu \nu } = 0$), but it is not traceless ($T ^{\mu} _{\phantom{\mu} \mu} \neq 0$). As a consequence, unlike most of the cases where Lorentz symmetry is preserved, it cannot be symmetrized \cite{Petrov1}.

In practice, the Casimir pressure can be computed following different methods. One possible route is the evaluation of the physical vacuum energy, which is defined as the difference between the zero-point energy in the presence of boundaries and that of the vacuum \cite{Schwinger1, Schwinger2, Heisenberg}. This definition only makes sense if it is combined with appropriate regularization methods which guarantee a finite expression for this energy difference; however, this method remains a problematic exercise, because standard regularization techniques, in most cases, only allow an approximate calculation \cite{Greiner}. Another procedure which allows the direct computation of the Casimir pressure is a field theoretical approach, where one examines the constrained propagation of virtual field quanta and considers the vacuum stress tensor, which can be expressed in terms of propagators \cite{Miltonbook, Brown, Deutsch}. Although both techniques can be shown to be formally equivalent, specific difficulties emerge in each case. For example, on the one hand, the mode-summation method requires the knowledge of the total energy spectrum of the free and the constrained field modes, besides we have to employ a particular cutoff  procedure to deal with divergences. On the other hand, the local approach requires the determination of exact Green functions describing propagation in the presence of external boundaries. In this paper we will employ the local formulation for the sake of simplicity, as shall see below.

From a quantum field theory approach, to study the properties of the vacuum requires the analysis of the behavior of local field quantities, such as the stress-energy tensor $T ^{\mu \nu}$. In the presence of a boundary $\Sigma$, the vacuum stress-energy tensor $\left< T ^{\mu \nu} \right>$ can be defined in the form
\begin{align}
\left< T ^{\mu \nu} \right> = \left< \, 0 \, \vert T ^{\mu \nu} \vert \, 0 \, \right> _{\Sigma} -  \left< \, 0 \, \vert T ^{\mu \nu} \vert \, 0 \, \right> _{0} ,  \label{vev-Tmunu}
\end{align}
i.e. the measurable quantities of the vacuum are defined as the difference between that in 
the constrained field configuration and the one corresponding to the unconstrained field. 
The space-space components of the vacuum stress (\ref{vev-Tmunu}) are useful to derive 
the mechanical properties of the system, such as the pressure.

In the problem at hand we have to compute the Casimir pressure upon a cylindrical shell $\mathcal{C}$, and hence, in the cylindrical coordinate system defined by ($\rho,\varphi,z$), the Casimir pressure $\mathcal{F}$ (Casimir force per unit area $F/A$) can be derived from the radial-radial component of the vacuum stress as
\begin{align}
\mathcal{F} = F/A =  \langle \, 0 \, \vert \, T ^{\rho \rho} \vert \, 0 \, \rangle _{ \mathcal{C} _{\mbox{\scriptsize in}} } - \langle \, 0 \, \vert \, T ^{\rho \rho} \, \vert \, 0 \, \rangle _{\mathcal{C} _{\mbox{\scriptsize out}} }   ,\label{F1a}
\end{align}
where $\mathcal{C} _{\mbox{\scriptsize out}}$  and $\mathcal{C} _{\mbox{\scriptsize in}}$ are the cylindrical surfaces immediately outside and inside the physical cylinder $\mathcal{C}$, respectively. It turns out that the vacuum stress is expressible in terms of field propagators, since the latter can be defined as the vacuum expectation value of the time-ordered product of two fields, i.e. \cite{Brown, Deutsch},
\begin{align}
G (x , x ^{\prime} ) = - i \,\langle\, 0 \, \vert \, \hat{\mathcal{T}} \phi(x) \phi( x ^\prime) \, \vert 0 \, \rangle .  \label{Grel}
\end{align}
In this way, the occurrence of quantum field fluctuations and the associated observable vacuum effects can be understood from the modifications in the propagation of virtual field quanta under external constraints. Indeed, the vacuum stress  formally follows after performing the Lorentz-covariant limit $x ^{\prime} \to x $. Substituting Eq. (\ref{Grel}) into the expression for the Casimir pressure (\ref{F1a}) we obtain
\begin{align}
\mathcal{F} = - \frac{i}{2} (1 + \lambda \xi ^{2}) \lim _{x ^{\prime} \rightarrow {x}}  \frac{\partial ^{2} }{\partial \rho \, \partial \rho ^{\prime}} \big[ G _{\mbox{\scriptsize in}} (x , x ^{\prime} ) -  G _{\mbox{\scriptsize out}} (x , x ^{\prime} ) \big] \bigg| _{ \rho = R }  , \label{FAbyG}
\end{align}
where $G _{\mbox{\scriptsize in}}$ and $G _{\mbox{\scriptsize out}}$ are the field propagators for the quantum fields inside and outside the shell, respectively, and $\xi = u ^{\mu} n _{\mu}$ is the projection of the background vector $u ^{\mu}$ along the outward unit normal to the cylinder $n _{\mu} = (0 , \vec{e} _{\rho})$. Clearly, a background of the form $u ^{\mu} = (1,0,0,0)$ leaves intact the form of the Casimir pressure. In general, there are other terms not included in Eq. (\ref{FAbyG}) which depend on angular derivatives of the GFs, and therefore they vanish by virtue of the Dirichlet BC on the surface. So, for the evaluation of the Casimir pressure, as dictated by the expression (\ref{FAbyG}), we require the Green's function for the quantum field both inside and outside the cylindrical shell. The next section is devoted to the detailed evaluation of these propagators.

\section{Green's function}\label{GFsection}

In order to evaluate the Casimir pressure given by Eq. (\ref{FAbyG}), in this section we derive analytical expressions for the field propagators inside and outside the shell, satisfying Dirichlet boundary conditions on the surface $\rho = R$. From Eq. (\ref{EQLV}) it follows that the corresponding GF satisfies the equation
\begin{align}
\left[ \Box  + \lambda \left( u _{\mu} \partial ^{\mu} \right) ^{2} +  m ^{2} \right] G( x , x ^{\prime}) = - \delta (x - x ^{\prime}) \ , \label{GF eq}
\end{align}
such that 
\begin{align}
G( x , x ^{\prime} ) \, \big| _{\rho = R} =  0 \  . \label{Dirichlet}
\end{align}
Also, we assume the physical requirement that $G$ be bounded as $\rho$ recedes to infinity, as well as at the origin. Due to the translational invariance in time, we can introduce the Green's function in the frequency domain $ G _{\omega} (\vec{x} , \vec{x} ^{\prime} )$ through the Fourier transformation
\begin{align}
G  ( x , x ^{\prime} ) = \frac{1}{2 \pi}\int _{- \infty} ^{\infty} d\omega \, e ^{i\, \omega (t-t ^{\prime})} G _{\omega}(\vec{x} , \vec{x} ^{\, \prime} ) \ . \label{GFT}
\end{align}
Here, $\vec{x}$ is the spatial part of the four-vector $x$, i.e. $x = ( t ,\vec{x} )$. Substituting Eq. (\ref{GFT}) into Eq. (\ref{GF eq}) we find that the GF in the frequency domain $G _{\omega} ( \vec{x} , \vec{x} ^{\, \prime} )$ satisfies the differential equation
\begin{align}
\left[ \omega ^{2} + \vec{\nabla} ^{2} - \lambda \, (i u _{0} \omega - \vec{u} \cdot \vec{\nabla} ) ^{2} - m ^{2} \right] G _{\omega} ( \vec{x} , \vec{x} ^{\, \prime} ) = \delta (\vec{x} - \vec{x} ^{\, \prime} ) \ . \label{GF eq omega}
\end{align}
The presence of the constant vector $\vec{u}$ breaks the usual rotational symmetry, and hence, to obtain explicit solutions to Eq. (\ref{GF eq omega}), we need to specify the form of $\vec{u}$. In the present study two particular cases will be analyzed in detail, namely
\begin{itemize}
  \item[(I)] the radial spacelike case $u ^{\mu} = (0,\vec{e} _{\rho})$, so, $u^\mu$ is an outward normal unit vector to the cylinder, and

  \item[(II)] the timelike case for which $u ^{\mu} = (1,\vec{0} \, )$, where the timelike component $u _{0}$ is different from zero only.
\end{itemize}
In the rest of this section we compute the Green's function for the two cases above, since they will fully determine the Casimir stress through Eq. (\ref{FAbyG}). 

\subsection{Radial spacelike case}

Taking $u ^{\mu} = (0,\vec{e} _{\rho})$, the differential equation (\ref{GF eq omega}) becomes
\begin{align}
\left[ \omega ^{2} + \frac{\partial^2}{\partial z ^{2} } + \Lambda \frac{\partial ^{2}}{\partial \rho ^2} + \frac{1}{\rho}\frac{\partial}{\partial \rho} + \frac{1}{\rho ^{2}} \frac{\partial ^{2}}{\partial\theta ^{2}} \right] G _{\omega} (\vec{x} _{\perp} , z ; \vec{x} _{\perp} ^{\, \prime} , z ^{\prime} )= - \delta (z-z ^{\prime}) \delta (\vec{x} _{\perp} - \vec{x} _{\perp} ^{\, \prime})
  \ , \label{GFLV ur}
\end{align}
where $\Lambda \equiv 1 - \lambda$ encodes the Lorentz-breaking contribution. To solve this equation, we use the fact that the delta function $\delta (z-z ^{\prime}) \delta (\vec{x} _{\perp} - \vec{x} _{\perp} ^{\, \prime})$ admits the factorization
\begin{align}
    \delta (z-z ^{\prime}) \delta (\vec{x} _{\perp} - \vec{x} _{\perp} ^{\, \prime}) = \frac{\delta (z-z ^{\prime}) \delta (\rho  - \rho ^{\prime}) \delta (\theta - \theta ^{\prime})}{\rho} = \frac{\delta (\rho  - \rho ^{\prime})}{\rho} \left( \frac{1}{2 \pi} \int _{- \infty} ^{+ \infty} e ^{i k (z-z ^{\prime})} dk \right) \left( \frac{1}{2 \pi} \sum _{m = - \infty} ^{+ \infty} e ^{im (\theta - \theta ^{\prime})} \right) . \label{completeness}
\end{align}
Accordingly, the GF we consider has translational invariance in the $z$ direction, while this invariance is broken in the $\rho$ direction. Also, it has rotational invariance around the axis of the cylinder. Therefore, we introduce the following 2+1 representation for the GF:
\begin{align}
\label{Greenansatz}
G _{\omega} (\vec{x} _{\perp} , z ; \vec{x} _{\perp} ^{\, \prime} , z ^{\prime} ) = \frac{1}{(2 \pi) ^{2}} \sum _{m = - \infty} ^{+ \infty} \int _{- \infty} ^{+ \infty} dk \, e ^{i k (z-z ^{\prime})} e ^{im (\theta - \theta    ^{\prime})} \, g _{m} (\rho , \rho ^{\prime} ; k ) \ ,
\end{align}
where $k$ is the momentum in the $z$ direction, and $g _{m} (\rho , \rho ^{\prime};k )$ is the reduced GF, which encodes all the information regarding the presence of boundaries with cylindrical symmetry. In the following, we suppress the dependence on $k$ of the reduced GF $g _{m}$. To go forward, we have to derive the differential equation for the reduced GF. To this end, we insert the 2+1 representation of Eq. (\ref{Greenansatz}) and the completeness relation (\ref{completeness}) into Eq. (\ref{GFLV ur}) to obtain
\begin{align}
\left[  \omega ^{2} - k ^{2} + \Lambda \, \frac{\partial ^2}{\partial \rho ^2} + \frac{1}{\rho}\frac{\partial}{\partial \rho} - \frac{m ^{2}}{\rho ^{2}} \right] g _{m} (\rho , \rho ^{\prime} ) = - \frac{ \delta (\rho  - \rho ^{\prime})}{\rho} \ . \label{ReducedGFEq}
\end{align}
Integrating twice this expression we obtain the corresponding BCs at $\rho = \rho ^{\prime}$, i.e.
\begin{align}\label{BC1}
    g _{m} (\rho , \rho ^{\prime} ) \Big| _{\rho = \rho ^{\, \prime} + 0 ^{+} } - g _{m} (\rho , \rho ^{\, \prime} ) \Big| _{\rho = \rho ^{\prime} -  0 ^{+} } &= 0 \ , \\ \frac{\partial g _{m} (\rho , \rho ^{\, \prime} )}{\partial \rho} \Bigg| _{\rho = \rho ^{\, \prime} + 0 ^{+} } -  \frac{\partial g _{m} (\rho , \rho ^{\, \prime} )}{\partial \rho} \Bigg| _{\rho = \rho ^{\, \prime} - 0 ^{+} } & =  - \frac{1}{\Lambda \rho ^{\, \prime}}  \  . \label{BC2}
\end{align}
Also, we impose the Dirichlet BC at the surface of the cylinder, namely $g _{m} (\rho = R , \rho ^{\prime} ) = 0$. Therefore, the differential equation (\ref{ReducedGFEq}) can be interpreted as consisting of the homogeneous equation for $\rho \in \mathbb{R} \setminus \{\ \!\! \rho ^{\prime} \}\ \!\! $, together with the boundary conditions at $\rho = \rho ^{\prime}$ and $\rho = R$. Indeed, the solution to the homogeneous equation is simple. It can be written as a linear combination of Bessel functions of the first and second kind as
\begin{align}\label{gmgen}
g _{m} (\rho , \rho ^{\, \prime} ) =  \rho ^{\frac{\Lambda -1}{2\,\Lambda }} \, \left[ c _{1} \, J _{n _{m}} (  \chi \rho  ) + c _{2} \, Y _{n _{m}} (  \chi \rho  ) \right] \ ,
\end{align}
where $c _{1}$ and $c _{2}$ are constants of integration and
\begin{align}
n _{m} = \sqrt{ \frac{m ^{2}}{ \Lambda } + \left( \frac{1 - \Lambda }{2 \Lambda } \right) ^{2} } , \qquad \chi = \sqrt{\frac{ \omega ^{2} - k ^{2}}{\Lambda }}\  .
\end{align}
Clearly, in the Lorentz-symmetric limit ($\lambda \to 0$) we obtain $n _{m} = m$ and $\chi = q$, as it should be. Now we can evaluate the GF both in the inner and the outer region.

%\subsection*{Inner region}

\begin{figure}
\includegraphics[scale=0.34]{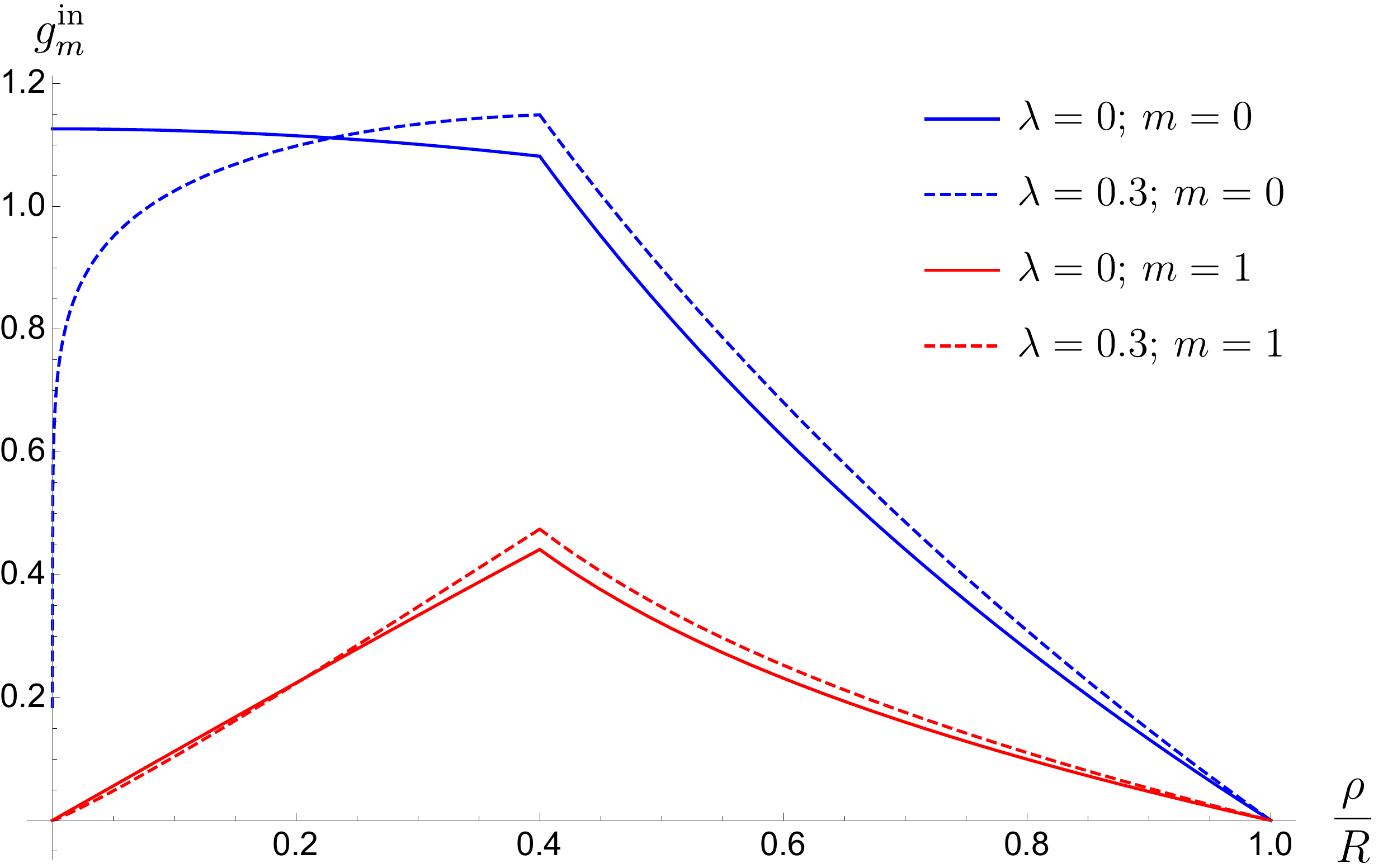}
\caption{The reduced GF $g _{m} ^{\mbox{\scriptsize in}} (\rho , \rho ^{\, \prime} )$ for a massless scalar field inside the cylinder  as a function of $\rho / R$, for $\sqrt{\omega ^{2} - k ^{2}} R = 1$ and $\rho ^{\prime} / R =0.4$. } \label{Plot-RedGF-inner}
\end{figure}

To compute the reduced GF in the inner region, we assume $\rho ^{\prime} <R$ and $\rho <R$. Finiteness of the GF at the origin implies that it can be taken as
\begin{align}
g _{m} ^{\mbox{\scriptsize in}} (\rho , \rho ^{\prime} ) = \rho ^{\frac{\Lambda - 1}{2\,\Lambda }}  \left\lbrace \begin{array}{l} A \, J _{n _{m}} ( \chi \rho  ) \\[7pt] B \, J _{n _{m}} ( \chi \rho  ) + C \, Y _{n _{m}} ( \chi \rho  ) \end{array} \begin{array}{l} \mbox{if } \rho < \rho ^{\, \prime} < R \\[7pt]  \mbox{if }  \rho ^{\, \prime} < \rho < R \end{array} \right. \ , \label{ansatz-inner}
\end{align}
since $Y _{n _{m}}$ diverges as the argument approaches to zero. The coefficients $A$, $B$ and $C$ can be determined by imposing the boundary conditions at $\rho = \rho ^{\prime}$ and at the surface of the cylinder. On the one hand, the Dirichlet condition at $\rho = R$ implies $B = - C \, Y _{n _{m}} ( \chi R  ) / J _{n _{m}} ( \chi R  )$, thus leaving us just with the unknown coefficients $A$ and $C$. The remaining BCs (\ref{BC1})-(\ref{BC2}) at $\rho = \rho ^{\prime}$ produce the system of algebraic equations
\begin{align}
A J _{n _{m}} ( \chi \rho ^{\prime} ) = C \left[ Y _{n _{m}} ( \chi \rho ^{\prime} ) -  \frac{Y _{n _{m}} ( \chi R )}{J _{n _{m}} ( \chi R )} \, J _{n _{m}} ( \chi \rho ^{\prime} ) \right] , \\ C \left[ \,Y _{n _{m}} ^{\prime} ( \chi \rho ^{\prime} ) - \frac{ Y _{n _{m}} ( \chi R ) }{J _{n _{m}} ( \chi R )}\,J _{n _{m}} ^{\prime}  ( \chi \rho ^{\prime} ) \,\right] - A J _{n _{m}} ^{\prime}  ( \chi \rho ^{\prime} )  = -  \frac{1}{\Lambda \chi} \frac{1}{\rho ^{\prime \, \frac{3\,\Lambda -1}{2\,\Lambda }} }\ ,
\end{align}
which can be easily solved to obtain
\begin{align}
C &= - \frac{\pi / 2}{\Lambda} \frac{1}{\rho ^{\prime \, \frac{\Lambda -1}{2\,\Lambda } } }   J _{n _{m}} ( \chi \rho ^{\prime} ) , \\ A &= - \frac{\pi / 2}{ \Lambda} \frac{1}{\rho ^{\prime \, \frac{\Lambda -1}{2\,\Lambda }} }  \left[ Y _{n _{m}} ( \chi \rho ^{\prime} ) -  \frac{Y_{n _{m}} ( \chi R )  }{J _{n _{m}} ( \chi R )}\,J _{n _{m}} ( \chi \rho ^{\prime} ) \right]  \ .
\end{align}
Upon substitution of the  coefficients into the ansatz (\ref{ansatz-inner}) we observe that the reduced GF can be written in the compact form
\begin{align}\label{gmin}
g _{m} ^{\mbox{\scriptsize in}} (\rho , \rho ^{\, \prime} ) = - \frac{\pi }{2\,\Lambda} \left( \frac{\rho}{\rho ^{\prime}} \right) ^{\frac{\Lambda -1}{2\,\Lambda }} \, \frac{J _{n _{m}} ( \chi \rho  _{<} )}{J _{n _{m}} ( \chi R )}\, \left[\, Y _{n _{m}} ( \chi \rho _{>} ) J _{n _{m}} ( \chi R ) \ - \ Y _{n _{m}} ( \chi R ) J _{n _{m}} ( \chi \rho _{>} )\, \right] \ ,
\end{align}
where $\rho _{>}$ ($\rho _{<}$) denotes the greater (smaller) value between $\rho$ and $\rho ^{\prime}$. In Fig. \ref{Plot-RedGF-inner} we plot the reduced GF of Eq. (\ref{gmin}) for a massless scalar field as a function of the dimensionless radius $\rho / R$ for $\sqrt{\omega ^{2} - k ^{2}} R = 1$, $\rho ^{\prime} / R =0.4$, and different values of $m$ and $\lambda$.

%\subsection*{Outer region}

\begin{figure}
\includegraphics[scale=0.5]{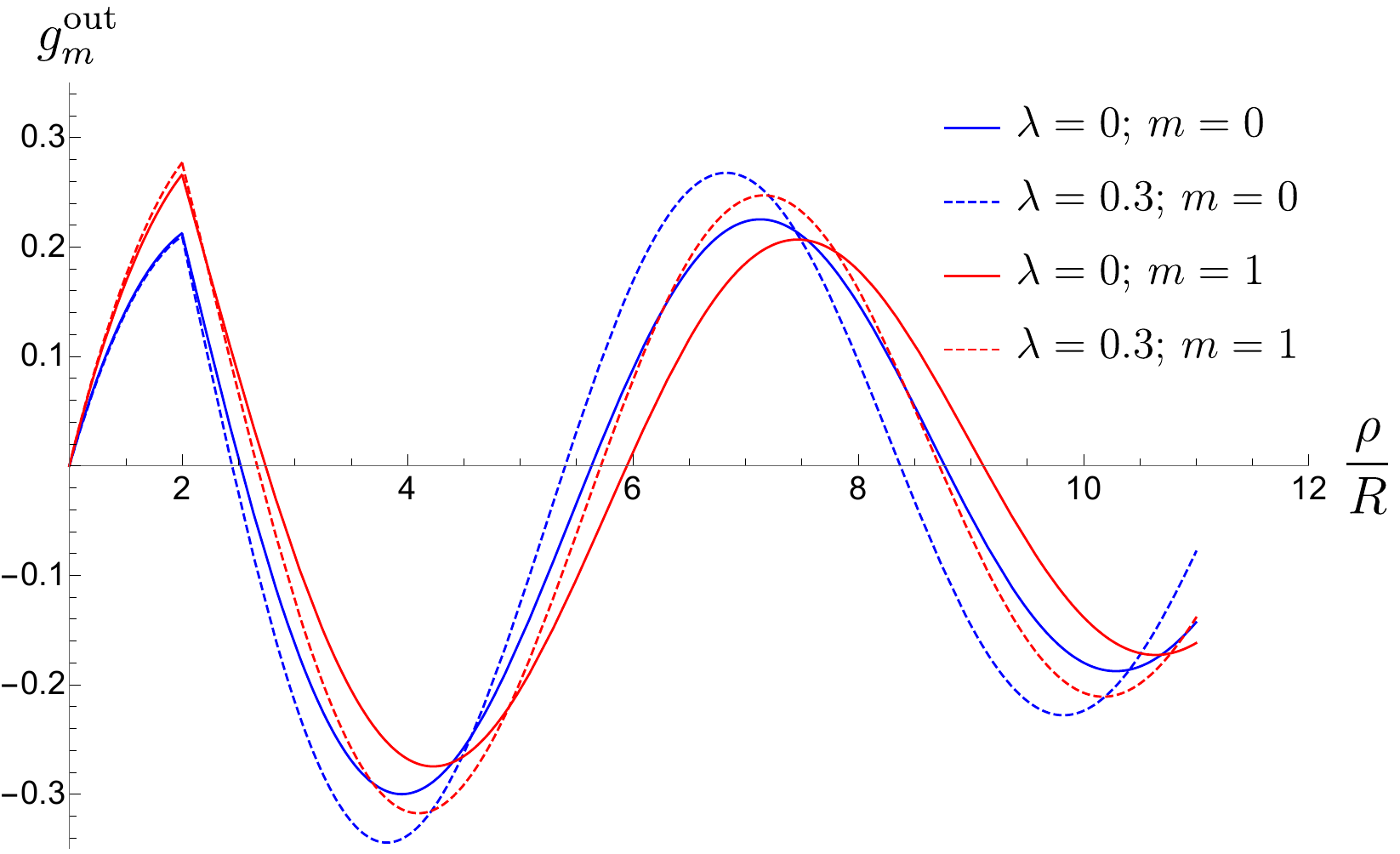}
\caption{The reduced GF $g _{m} ^{\mbox{\scriptsize out}} (\rho , \rho ^{\, \prime} )$ for a massless scalar field outside the cylinder as a function of $\rho / R$, for $\sqrt{\omega ^{2} - k ^{2}} R = 1$ and $\rho ^{\prime} / R =2$. } \label{Plot-RedGF-outer}
\end{figure}

To compute GF outside the cylinder, it is convenient to express the solution (\ref{gmgen}) in terms of the Hankel functions of the first and second kind as
\begin{align}
g _{m} ^{\mbox{\scriptsize out}} (\rho , \rho ^{\, \prime} ) = \rho ^{\frac{\Lambda -1}{2\,\Lambda }} \left\lbrace \begin{array}{l} A \, H _{n _{m}} ^{(1)} ( \chi \rho ) + B \, H _{n _{m}} ^{(2)} ( \chi \rho ) \\[7pt] C \, H _{n _{m}} ^{(1)} ( \chi \rho ) \end{array} \begin{array}{l} \mbox{if } R< \rho < \rho ^{\, \prime} \\[7pt]  \mbox{if } R <  \rho ^{\, \prime} < \rho \end{array} \right. \ .
\end{align}
Imposing the Dirichlet condition at $\rho = R$ we obtain $ B = - A \, H _{n _{m}} ^{(1)} ( \chi R ) / H _{n _{m}} ^{(2)} ( \chi R )$, thus leaving us with the undetermined coefficients $A$ and $C$. Next, the two BCs (\ref{BC1})-(\ref{BC2}) at $\rho=\rho^\prime$ fix the values of $A$ and $C$:
\begin{align}
A &= \frac{\pi}{4 i} \frac{1}{\Lambda} \frac{1}{\rho ^{\prime \, {\frac{\Lambda -1}{2\,\Lambda }}}} \, H _{n _{m}} ^{(1)} ( \chi \rho ^{\prime} )  \frac{H _{n _{m}} ^{(2)} ( \chi R ) }{H _{n _{m}} ^{(1)} ( \chi R )} , \\ C & = \frac{\pi}{4 i} \frac{1}{\Lambda} \frac{1}{\rho ^{\prime \, {\frac{\Lambda -1}{2\,\Lambda }}}}  \frac{H _{n _{m}} ^{(2)} ( \chi R ) }{H _{n _{m}} ^{(1)} ( \chi R )}  \left[ H _{n _{m}} ^{(1)} ( \chi \rho ^{\prime} ) - \frac{H _{n _{m}} ^{(1)} ( \chi R ) }{H _{n _{m}} ^{(2)} ( \chi R )}  \, H _{n _{m}} ^{(2)} ( \chi \rho ^{\prime} ) \right] \ .
\end{align}
Eventually, we obtain the compact solution
\begin{align}\label{gmout}
g _{m} ^{\mbox{\scriptsize out}} (\rho , \rho ^{\, \prime} ) = \frac{\pi}{4 \,i} \frac{1}{\Lambda} \left( \frac{\rho}{\rho ^{\prime}} \right) ^{\frac{\Lambda -1}{2\,\Lambda }}\, \frac{H _{n _{m}} ^{(1)} ( \chi \rho _{>} )}{H _{n _{m}} ^{(1)} ( \chi R )}  \, \left[ H _{n _{m}} ^{(1)} ( \chi \rho _{<} ) H _{n _{m}} ^{(2)} ( \chi R )\ - \ H _{n _{m}} ^{(1)} ( \chi R )  \, H _{n _{m}} ^{(2)} ( \chi \rho _{<} ) \right] \ ,
\end{align}
where, as before, $\rho _{>}$ ($\rho _{<}$) denotes the greater (smaller) value between $\rho$ and $\rho ^{\prime}$. In Fig. \ref{Plot-RedGF-outer} we plot the reduced GF of Eq. (\ref{gmout}) for a massless scalar field as a function of the dimensionless radius $\rho / R$ for $\sqrt{\omega ^{2} - k ^{2}} R = 1$, $\rho ^{\prime} / R =2$, and different values of $m$ and $\lambda$.

\subsection{Timelike case}
\label{tmG1}

Now we shall consider that the Lorentz-violating background tensor is given by the four vector $u^{\mu}=(1,\vec{0})$. In this case, as can be seen in Eq. (\ref{GF eq omega}), the Green's function satisfies
\begin{align}
\left[ (1+\lambda)\omega ^{2} + \frac{\partial^2}{\partial z^2} + \frac{\partial ^2}{\partial \rho ^2} + \frac{1}{\rho}\frac{\partial}{\partial \rho} +  \frac{1}{\rho^2}\frac{\partial^2}{\partial\theta^2} \right] G _{\omega} (\vec{x} _{\perp} , z ; \vec{x} _{\perp} ^{\, \prime} , z ^{\prime} ) = - \delta (z-z ^{\prime}) \,\delta (\vec{x} _{\perp} - \vec{x} _{\perp}^{\, \prime})
  \ , \label{GFLV ur}
\end{align}
which upon the redefinition $\omega ^{\prime} = \sqrt{1 + \lambda} \, \omega$, corresponds to the standard Lorentz-symmetric case (previously studied in Ref. \cite{Miltonbook}). Indeed, in this case the reduced GFs in the inner and outer regions are still given by equations (\ref{gmin}) and (\ref{gmout}), respectively, with $\Lambda = 1$ and $\omega \to \omega ^{\prime} = \sqrt{1 + \lambda} \, \omega$. Since the contribution is trivial, from now on we will primarily focus on the radial spacelike case.

\section{Casimir effect}\label{CEsection}

This section is devoted to the evaluation of the Casimir pressure, whose general expression is given by Eq. (\ref{FAbyG}). Substituting the Green's functions for the inner and outer regions, given by Eq. (\ref{GFT}) with the reduced GFs $g _{m} ^{\mbox{\scriptsize in}} (\rho , \rho ^{\, \prime} )$ and $g _{m} ^{\mbox{\scriptsize out}} (\rho , \rho ^{\, \prime} )$ computed in the previous section, we obtain the expression for the Casimir pressure:
\begin{align}
\mathcal{F} = - \frac{i (1 + \lambda \xi ^{2}) }{16 \pi ^{3}}  \sum _{m = - \infty} ^{+ \infty} \int _{- \infty} ^{+ \infty} d \omega \int _{- \infty} ^{+ \infty} dk \,  \lim _{\rho ^{\prime} \to \rho}   \frac{\partial ^{2}}{\partial \rho \partial \rho ^{\prime}} \left[ g _{m} ^{\mbox{\scriptsize in}} (\rho , \rho ^{\prime}) - g _{m} ^{\mbox{\scriptsize out}} (\rho , \rho ^{\prime}) \right] \bigg| _{\rho = R} . \label{CasPressure}
\end{align}
We observe that any term containing at least one derivative with respect to the angle $\theta$ or to the coordinate $z$ vanishes by virtue of the Dirichlet boundary condition at $\rho = R$, since the reduced GFs are zero there. This is why in Eq. (\ref{FAbyG}) we disregard additional terms. Having determined the above general expression for the Casimir pressure in terms of the reduced GF, we can now obtain explicit formulas for each case.

\subsection{Radial spacelike case}

Now we have to evaluate the derivatives required by Eq. (\ref{CasPressure}) and then compute the limit $\rho ^{\prime} \to \rho$. To this end, without loss of generality we take $\rho > \rho ^{\prime}$ in Eq. (\ref{gmin}) for the inner region, from which we obtain
\begin{align}
\lim _{\rho ^{\prime} \to \rho}  \, \frac{\partial ^{2}}{\partial \rho \partial \rho ^{\prime}}  \, g _{m} ^{\mbox{\scriptsize in}}  (\rho , \rho ^{\prime} ) \bigg| _{\rho = R} & =   - \frac{1}{\Lambda R ^{2}} \left[ \,\chi R \, \frac{J _{n _{m}} ^{\prime} ( \chi R )}{J _{n _{m}} ( \chi R )} - \frac{\Lambda-1}{2\,\Lambda } \,\right] \ ,
\end{align}
where the prime in the Bessel function denotes derivative with respect to its argument. A similar procedure for the exterior region (for which we take $R < \rho < \rho ^{\prime}$) yields
\begin{align}
\lim _{\rho ^{\prime} \to \rho} \,  \frac{\partial ^{2}}{\partial \rho \partial \rho ^{\prime}}  \, g _{m} ^{\mbox{\scriptsize out}}  (\rho , \rho ^{\prime} ) \bigg| _{\rho = R} & = \frac{1}{ \Lambda R ^{2}}  \left[ \,\chi R \frac{H _{n _{m}} ^{(1) \prime} ( \chi R )}{H _{n _{m}} ^{(1)} ( \chi R )} - \frac{\Lambda-1}{2\,\Lambda} \, \right] \ .
\end{align}
Finally, inserting these results into Eq. (\ref{CasPressure}) and noticing that $\xi ^{2} = -1$ there, we obtain the following expression for the Casimir pressure:
\begin{align}
\mathcal{F} = \frac{i}{4 \pi ^{3} R ^{2}} \sum _{m = - \infty} ^{+ \infty}  \int _{0} ^{\infty} d \omega \int _{0} ^{\infty} dk \, \left\lbrace  \chi R \, \frac{J _{n _{m}} ^{\prime} ( \chi R )}{J _{n _{m}} ( \chi R )} +  \chi R \frac{H _{n _{m}} ^{(1) \prime} ( \chi R )}{H _{n _{m}} ^{(1)} ( \chi R )}  -  \frac{\Lambda-1}{\Lambda } \right\rbrace  \  .
\label{FAr}
\end{align}
The frequency integral is best done by performing a complex frequency rotation $\omega = i \zeta $. In this case, the above Casimir force (\ref{FAr}) can be expressed in terms of the modified Bessel functions of the first and second kind, namely
\begin{align}
\mathcal{F} = - \frac{1}{4 \pi ^{3} R ^{2}} \sum _{m = - \infty} ^{+ \infty}  \int _{0} ^{\infty} d \zeta \int _{0} ^{\infty} dk \, \left\lbrace \chi ^{\ast} R \frac{I _{n _{m}} ^{\prime} ( \chi ^{\ast}  R )}{I _{n _{m}} (  \chi ^{\ast}  R )} + \chi ^{\ast} R \frac{K _{n _{m}} ^{\prime} ( \chi ^{\ast}  R )}{K _{n _{m}} ( \chi ^{\ast} R )} - \frac{\Lambda-1}{\Lambda} \right\rbrace \ ,  \label{FArcom}
\end{align}
where $ \chi ^{\ast} = \sqrt{ ( \zeta ^{2} + k ^{2} ) / \Lambda }$. Now making the change of variables $\zeta = \sqrt{\Lambda} \frac{\Delta}{R} \cos \alpha$ and $k = \sqrt{\Lambda} \frac{\Delta}{R} \sin \alpha$, where $\Delta \in [0, \infty )$ and $\alpha \in [0 , \pi /2]$, the angular integration is trivial and, after some algebraic simplifications, we obtain (restoring $\hbar$ and $c$)
\begin{align}
\mathcal{F} = - \Lambda \frac{\hbar c }{8 \pi ^{2} R ^{4}} \sum _{m = - \infty} ^{+ \infty}  \int _{0} ^{\infty} d \Delta  \, \Delta^2 \left\lbrace  \frac{d}{d\Delta} \ln \left[ I _{n_{m}}(\Delta) K _{n_{m}}(\Delta) \Delta^{ -  \frac{\Lambda-1}{\Lambda}} \right] \right\rbrace , \label{F1}
\end{align}
being $\Delta = R \sqrt{\frac{(\zeta / c) ^{2} + k ^{2}}{\Lambda}}$ a dimensionless variable of integration. With the help of this formula, in the next section we compute the Casimir pressure for the radial spacelike case.

\subsection{Timelike case}

This case is rather simpler. On the one hand, the factor $1 + \lambda \xi ^{2}$ in front of the expression for the Casimir pressure (\ref{CasPressure}) equals to unity, since $\xi$ is the projection of the Lorentz-violating tensor $u ^{\mu}$ along the normal to the cylinder $n ^{\mu} = (0, \vec{e} _{\rho})$, and hence $\xi = 0$. So, equation (\ref{CasPressure}) is the same as in the Lorentz symmetric case. On the other hand, as pointed out in Sec. \ref{tmG1}, the Green's function for the timelike case is the same as that for the Lorentz-symmetric GF, with the replacement $\omega \to \sqrt{1 + \lambda } \omega$ only. Therefore, the expression for the Casimir pressure for the timelike case can be obtained from that of Eq. (\ref{F1}) by taking $\Lambda = 1$ and replacing $\omega \to \sqrt{ \Lambda ^{\prime} } \omega$, with  $\Lambda ^{\prime} = 1 + \lambda$. The result is then
\begin{align}
\mathcal{F} _{t} &= - \frac{\hbar c}{8 \pi ^{2} R ^{4}}  \sum _{m = - \infty} ^{\infty} \int _{0} ^{\infty} d \Delta  \, \Delta ^{2} \,  \frac{d}{d\Delta} \ln \left[ I _{m}(\Delta) K _{m}(\Delta) \right]  \ ,  \label{Ft}
\end{align}
where now, the dimensionless variable of integration reads $\Delta = R \sqrt{(\zeta \Lambda ^{\prime} / c) ^{2} + k ^{2}}$.

The expressions (\ref{F1}) and (\ref{Ft}) for the Casimir pressure for the radial spacelike case and the timelike case, respectively, have infinite values, and as usual, it must be regularized in some way. In the next section, we use the zeta function regularization technique \cite{Elizalde1, Elizalde2} to get finite results for the expressions (\ref{F1}) and (\ref{Ft}).

\section{Towards numerical evaluation of the Casimir pressure}
\label{NR1}

\begin{figure}
\includegraphics[scale=0.35]{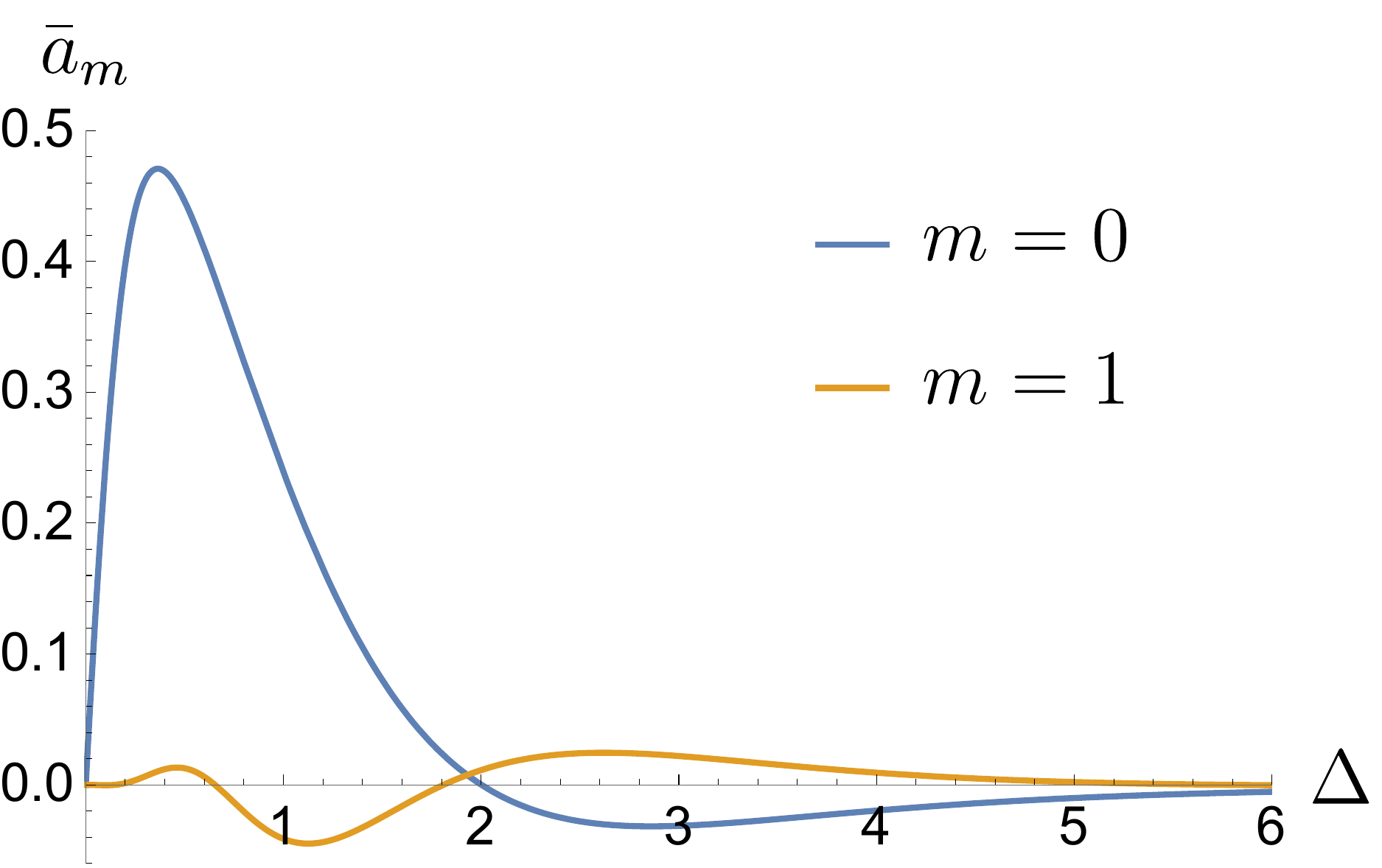}\hspace{0.45cm} \includegraphics[scale=0.35]{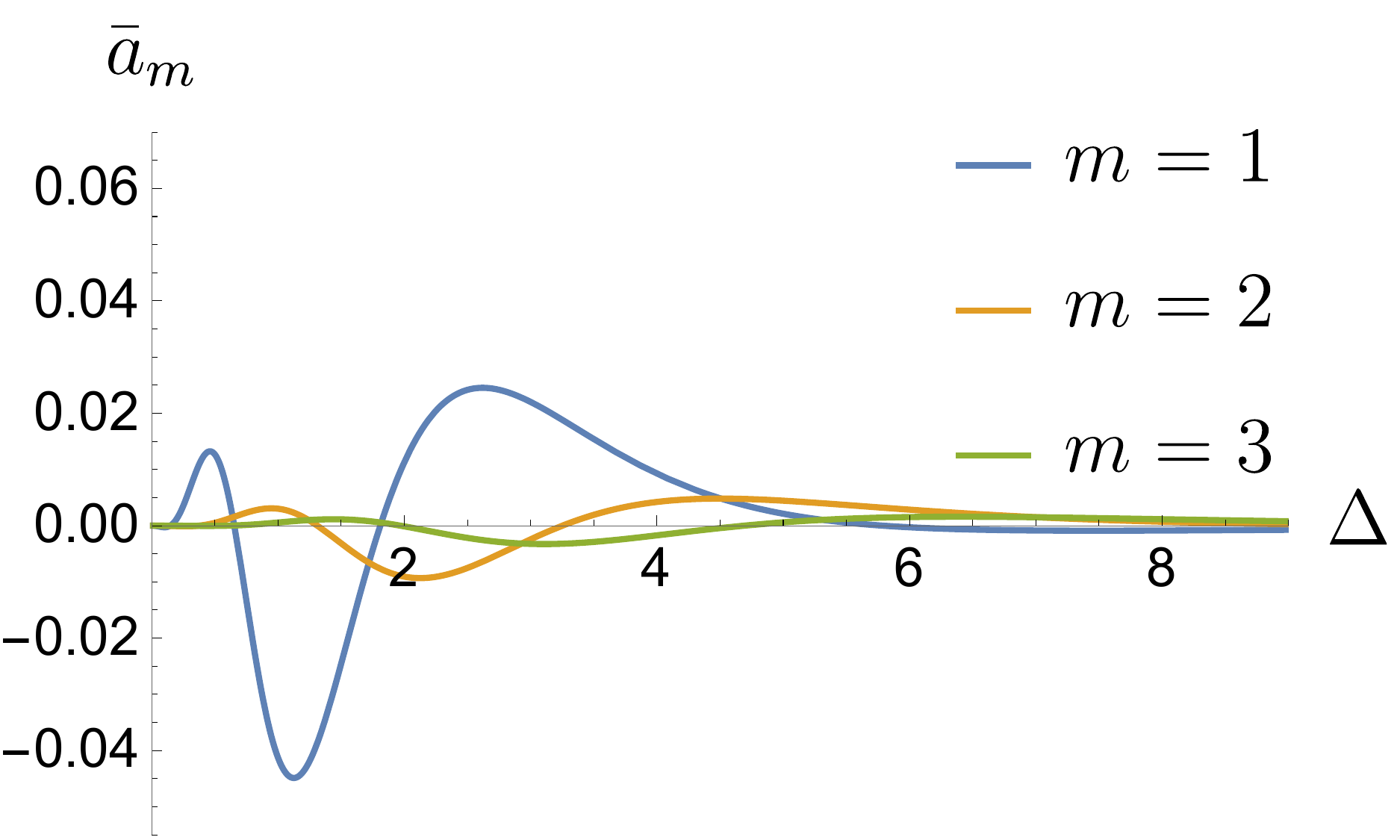}
\caption{The oscillatory integrand $\overline{a} _{m}(\Delta)$ in Eq. (\ref{CoeffAm}) rapidly decreases to zero as a function of $\Delta$ for any $m$, thus ensuring the convergence of the integrals. At fixed $m$, $\overline{A}_{m}>\overline{A}_{m+1}$ and the whole series in (\ref{f-regularize3}) is convergent. The above plots correspond to $\Lambda=1$, they exhibit the generic behavior for $\Lambda \lesssim 1$. } \label{AmB}
\end{figure}

In the original parallel plate configuration \cite{Casimir}, the Casimir pressure for a scalar field is $\mathcal{F} = - \frac{\pi ^{2}}{480}  \frac{\hbar c}{L ^{4}} $, which clearly depends only on the fundamental constants and the distance between the plates $L$. To obtain this finite result, different regularization techniques haven been used, for example, dimensional regularization through the Schwinger proper time representation of the vacuum energy \cite{Schwinger3} and the local zeta function regularization \cite{Elizalde1, Elizalde2}. The goal of this section is to evaluate the Casimir pressure as given by Eqs. (\ref{F1}) and (\ref{Ft}), for a perfectly conducting cylindrical shell. For the sake of simplicity, we discuss in detail the radial spacelike case, which as discussed above, is more intricate than the timelike case. To this end, from Eq. (\ref{F1}), we define the dimensionless quantity
\begin{align}
f ( \Lambda ) = \frac{\mathcal{F}}{\mathcal{F} _{0}} = \sum _{m = - \infty} ^{\infty} \mathcal{A} _{m} (\Lambda ) , \label{Freduced}
\end{align}
where 
\begin{align}
\mathcal{F} _{0} = - \frac{\hbar c}{8 \pi ^{2} R ^{4}}  \label{F0}
\end{align}
and
\begin{align}
\mathcal{A} _{m} (\Lambda) = \Lambda \int _{0} ^{\infty} d \Delta  \, \Delta ^{2}  \frac{d}{d\Delta} \textrm{ln} \left[ I _{n_{m}}(\Delta) K _{n_{m}}(\Delta) \Delta^{-  \frac{\Lambda-1}{\Lambda}} \right] .  \label{IntegralAm}
\end{align}
Although simple, the above expression is still not useful to evaluate $f(\Lambda )$ since the series, as well as the integral, do not converge. In order to obtain a convergent reformulation, we first look for the asymptotic behavior of the integrand of the function $\mathcal{A} _{m}$ as $n _{m} \gg 1$. To this end, we employ the uniform asymptotic expansions of the modified Bessel functions for large $\mu$ \cite{Abramowitz}: 
\begin{align}
I _{\mu} (\mu z) & \sim \frac{e ^{\mu \eta }}{\sqrt{2 \pi \mu } (1 + z ^{2})^{1/4}} \sum _{k = 0} ^{\infty} \frac{U _{k} (t)}{\mu ^{k}} ,  \notag \\ K _{\mu} (\mu z) & \sim \sqrt{\frac{\pi}{2 \mu }} \frac{e ^{- \mu \eta }}{(1 + z ^{2})^{1/4}} \sum _{k = 0} ^{\infty} (-1) ^{k} \frac{U _{k} (t)}{\mu ^{k}} ,  \label{AsymptoticBessels}
\end{align}
where
\begin{align}
\eta = (1+z ^{2}) ^{1/2} + \ln \frac{z}{1 + \sqrt{1 + z ^{2}}} , \qquad t = (1+z ^{2}) ^{- 1/2} ,
\end{align}
and 
\begin{align}
U _{0}(t) = 1 , \qquad U _{1} (t) = \frac{1}{24} (3t-5t ^{3}) , \qquad U _{2} (t) = \frac{1}{1152} (81 t ^{2} - 462 t ^{4} + 385 t ^{6}) .
\end{align}
With the above results, it is easy to see that
\begin{align}
\lim _{n _{m} \gg 1 } \Lambda \Delta ^{2}  \frac{d}{d\Delta} \textrm{ln} \left[ I _{n_{m}}(\Delta) K _{n_{m}}(\Delta) \Delta^{-  \frac{\Lambda - 1}{\Lambda}} \right] & \sim (1 - 2 \Lambda )  \Delta +  \frac{\Lambda n_{m} ^{2} \Delta }{n_{m} ^{2} + \Delta ^{2} } \notag \\ & \phantom{==} - \frac{2 \Lambda  \Delta ^{3} \left(4 n_{m} ^{4} - 10 n_{m} ^{2} \Delta ^{2} + \Delta  ^{4} \right)}{\left( n_{m} ^{2} + \Delta ^{2} \right) \left( 8 n_{m} ^{6} + 24 n_{m} ^{4} \Delta ^{2} + 4 n_{m} ^{2} \Delta ^{2} \left(6 \Delta  ^{2} - 1 \right) + 8 \Delta  ^{6} + \Delta  ^{4} \right)} ,
\end{align}
such that the asymptotic behavior of the coefficients (\ref{IntegralAm}) for $n _{m} \gg 1$ can be expressed as the sum of three divergent terms:
\begin{align}
\mathcal{A} _{m} ^{\mbox{\scriptsize Asymptot}} (\Lambda ) = \mathcal{B} _{m} ^{(1)} (\Lambda ) + \mathcal{B} _{m} ^{(2)}  (\Lambda ) + \mathcal{B} _{m} ^{(3)}  (\Lambda ) , \label{A-infinity}
\end{align}
where 
\begin{align}
\mathcal{B} _{m} ^{(1)} (\Lambda ) = \int _{0} ^{\infty} (1 - 2 \Lambda )  \Delta \, d \Delta , \qquad \mathcal{B} _{m} ^{(2)} (\Lambda ) = \int _{0} ^{\infty} \frac{\Lambda n _{m} ^{2} \Delta }{n_{m} ^{2} + \Delta ^{2} } \, d \Delta , \\ \mathcal{B} _{m} ^{(3)} (\Lambda ) = -  \int _{0} ^{\infty} \frac{2 \Lambda  \Delta ^{3} \left(4 n_{m} ^{4} - 10 n_{m} ^{2} \Delta ^{2} + \Delta  ^{4} \right)}{\left( n_{m} ^{2} + \Delta ^{2} \right) \left( 8 n_{m} ^{6} + 8 \Delta  ^{6} + \Delta  ^{4}  + 4 n_{m} ^{2} \Delta ^{2} \left(6 n_{m} ^{2} + 6 \Delta  ^{2} - 1 \right) \right)}  \, d \Delta . \label{B3}
\end{align}
Disregarding for the moment that the integrals in Eq. (\ref{B3}) diverge, we employ here the Riemann zeta function technique \cite{Elizalde1, Elizalde2} for attributing a finite value to the sum in Eq. (\ref{Freduced}):
\begin{align}
f (\Lambda ) &= \sum _{m = - \infty} ^{\infty} \left[ \mathcal{A} _{m} (\Lambda )  -  \mathcal{A} _{m} ^{\mbox{\scriptsize Asymptot}} (\Lambda )  \right] + \sum _{m = - \infty} ^{\infty}  \mathcal{A} _{m} ^{\mbox{\scriptsize Asymptot}} (\Lambda )  \notag \\ &= \sum _{m = - \infty} ^{\infty}  \overline{\mathcal{A}} _{m} (\Lambda )   + \sum _{m = - \infty} ^{\infty}  \mathcal{A} _{m} ^{\mbox{\scriptsize Asymptot}} (\Lambda )  , \label{f-regularize2}
\end{align}
where $\overline{\mathcal{A}} _{m} (\Lambda )$ stands for the renormalized part of the function, which can be explicitly written as
\begin{align}
\overline{\mathcal{A}} _{m} &= \int _{0} ^{\infty} d \Delta  \, \left\lbrace \Lambda   \Delta ^{2}  \frac{d}{d\Delta} \textrm{ln} \left[ I _{n_{m}}(\Delta) K _{n_{m}}(\Delta)  \Delta^{-  \frac{\Lambda-1}{\Lambda}}  \right] - \mathcal{A} _{m} ^{\mbox{\scriptsize Asymptot}} (\Lambda ) \right\rbrace . \label{CoeffAm}
\end{align}
We will back to this expression later. Now we focus in the second term of Eq. (\ref{f-regularize2}), which contains three divergent expressions. The first term, $\sum _{m = - \infty} ^{+ \infty}  \mathcal{B} _{m} ^{(1)} (\Lambda )$, contains the product of two divergent expressions, and we deal with more precisely by presenting it in the following form:
\begin{align}
\sum _{m = - \infty} ^{+ \infty}  \mathcal{B} _{m} ^{(1)} (\Lambda ) = (1 - 2 \Lambda )  \lim _{s \to 0 ^{+}} [2 \zeta (s) + 1 ] \int _{0} ^{\infty}  d \Delta \;\Delta \, e ^{- \sqrt{s} \Delta} ,
\end{align}
where $ \zeta (z)$ is the Riemann zeta function. Upon computing this integral we obtain
\begin{align}
\sum _{m = - \infty} ^{+ \infty}  \mathcal{B} _{m} ^{(1)} (\Lambda ) = (1 - 2 \Lambda )  \lim _{s \to 0 ^{+}}  \frac{2 \zeta (s) + 1}{s} \approx (1 - 2 \Lambda )  \lim _{s \to 0 ^{+}}  \frac{2 s \, \zeta ^{\prime} (s)}{s} = \left( 2 \Lambda - 1 \right)  \ln (2 \pi ) .
\end{align}
For the second term, $\sum _{m = - \infty} ^{+ \infty}  \mathcal{B} _{n} ^{(2)} (\Lambda )$ , we have
\begin{align}
\sum _{m = - \infty} ^{+ \infty}  \mathcal{B} _{m} ^{(2)} (\Lambda ) =  \Lambda \sum _{m = - \infty} ^{\infty} n _{m} ^{2} \int _{0} ^{\infty} d \Delta \, \frac{\Delta }{1 + \Delta ^{2}} = \Lambda \sum _{m = - \infty} ^{\infty} \left[ \frac{4m ^{2} \Lambda + (1 - \Lambda) ^{2}}{4 \Lambda ^{2}} \right] \int _{0} ^{\infty} d \Delta \, \frac{\Delta }{1 + \Delta ^{2}}
\end{align}
where the change variables $ \Delta \to n _{m} \Delta $ have been performed. The first part of the summation is simple: $\sum _{m = - \infty} ^{+ \infty}  m ^{2} = 2 \sum _{m = 1} ^{+ \infty}  m ^{2} =2 \zeta (-2) = 0$, since $\zeta (- 2 n) = 0$ for $n \in \mathbb{Z} ^{+}$ \cite{Abramowitz}. The second part of the summation contains a divergence, and we deal with by representing it in the following form:
\begin{align}
\sum _{m = - \infty} ^{+ \infty}  \mathcal{B} _{m} ^{(2)} (\Lambda ) &= \frac{(1 - \Lambda) ^{2}}{4 \Lambda }  \lim _{s \to 0 ^{+}} [2 \zeta (s) + 1 ]  \int _{0} ^{\infty} d \Delta \, \frac{\Delta }{1 + \Delta ^{2}} \, e ^{- \sqrt{s} \Delta} \notag \\ &= \frac{(1 - \Lambda) ^{2}}{4 \Lambda }  \lim _{s \to 0 ^{+}} [2 \zeta (s) + 1 ] \, \left[ - \cos (s) \mbox{Ci} (s) + \frac{1}{2} \sin (s) \left( \pi - 2 \mbox{Si} (s) \right) \right] ,
\end{align}
where $\mbox{Si}$ and $\mbox{Ci}$ are the sine and cosine integrals. Power expanding the above result we find
\begin{align}
\sum _{m = - \infty} ^{+ \infty}  \mathcal{B} _{m} ^{(2)} (\Lambda ) &=  \frac{(1 - \Lambda) ^{2}}{4 \Lambda }  \lim _{s \to 0 ^{+}} \ln (2 \pi ) \left[ \gamma + \ln (s) \right] s = 0 ,
\end{align}
where $\gamma$ is the Euler constant. Now we are left with the third divergent term $\sum _{m = - \infty} ^{+ \infty}  \mathcal{B} _{m} ^{(3)} (\Lambda )$, which is something different. On the one hand, we observe that the value of the integral is strongly dominated by an $m$-independent term of the form
\begin{align}
\max [ \mathcal{B} _{m} ^{(3)} ( \Lambda ) ] = \mathcal{B} _{0} ^{(3)} ( \Lambda ) = - \Lambda \int _{0} ^{\infty} \frac{2 \Delta }{ 1 + 8 \Delta  ^{2}}  \, d \Delta \approx -67, 760.93 \, \Lambda . 
\end{align}
The value of $\mathcal{B} _{m} ^{(3)} ( \Lambda )$ for $m \geq 1$ differ from that of $\mathcal{B} _{0} ^{(3)} ( \Lambda )$ by $87 \times 10 ^{-6}$ (for $m = 1$) and $19 \times 10 ^{-6}$ (for $m = 1 \times 10 ^{9}$). This means that, up to a precision of $10 ^{-6}$, the value of the integral is independent of $m$. This leaves us with a divergent expression, and we deal with more precisely by presenting it in the form
\begin{align}
\sum _{m = - \infty} ^{+ \infty}  \mathcal{B} _{m} ^{(3)} (\Lambda ) &= - \Lambda  \lim _{s \to 0 ^{+}} [2 \zeta (s) + 1 ] \int _{0} ^{\infty}  d \Delta \; \frac{2 \Delta }{ 1 + 8 \Delta  ^{2}} \, e ^{- \sqrt{s} \Delta} . 
\end{align}
Computing this integral we get
\begin{align}
\sum _{m = - \infty} ^{+ \infty}  \mathcal{B} _{m} ^{(3)} (\Lambda ) &= - \frac{\Lambda }{8}  \lim _{s \to 0 ^{+}} [2 \zeta (s) + 1 ]  \left\lbrace \left[ \pi -2 \text{Si} \left( \sqrt{s/8} \right) \right] \sin \left( \sqrt{s/8}  \right)-2 \text{Ci}\left( \sqrt{s/8}  \right) \cos \left( \sqrt{s/8}  \right) \right\rbrace , 
\end{align}
and power expanding the result we obtain
\begin{align}
\sum _{m = - \infty} ^{+ \infty}  \mathcal{B} _{m} ^{(3)} (\Lambda ) &= - \frac{\Lambda }{8}  \lim _{s \to 0 ^{+}} s \left\lbrace  \log (s) - \log \left( 8 \right)  + 2 \gamma   \right\rbrace \log (2 \pi ) = 0 . 
\end{align}
All in all, the expression (\ref{f-regularize2}) can thus be written in the form
\begin{align}
f (\Lambda ) &= \overline{\mathcal{A}} _{0} (\Lambda ) + 2  \sum _{m = 1} ^{\infty}  \overline{\mathcal{A}} _{m} (\Lambda )   + \left( 2 \Lambda - 1 \right)  \ln (2 \pi ) , \label{f-regularize3}
\end{align}
where we have used that $\overline{\mathcal{A}} _{m} = \overline{\mathcal{A}} _{- m}$. Clearly, this is a convergent expression for the Casimir pressure, which we now evaluate numerically. Firstly, let us define the integrand appearing in (\ref{CoeffAm}) as $\overline{a}_{m}$, namely
\begin{equation}\label{amreg}
\overline{a} _{m}(\Delta)  \equiv  \Lambda   \Delta ^{2}  \frac{d}{d\Delta} \textrm{ln} \left[ I _{n_{m}}(\Delta) K _{n_{m}}(\Delta)  \Delta^{-  \frac{\Lambda-1}{\Lambda}}  \right] - \mathcal{A} _{m} ^{\mbox{\scriptsize Asymptot}} (\Lambda )  .
\end{equation}
In Fig. \ref{AmB}, for the lowest values of $m=0,1,2,3$ and $\Lambda=1$ the behavior of $\overline{a} _{m}(\Delta)$ is shown. It manifests the correctness of the regularization procedure based on the Riemann zeta function technique.

\clearpage

\begin{figure}
\includegraphics[scale=0.4]{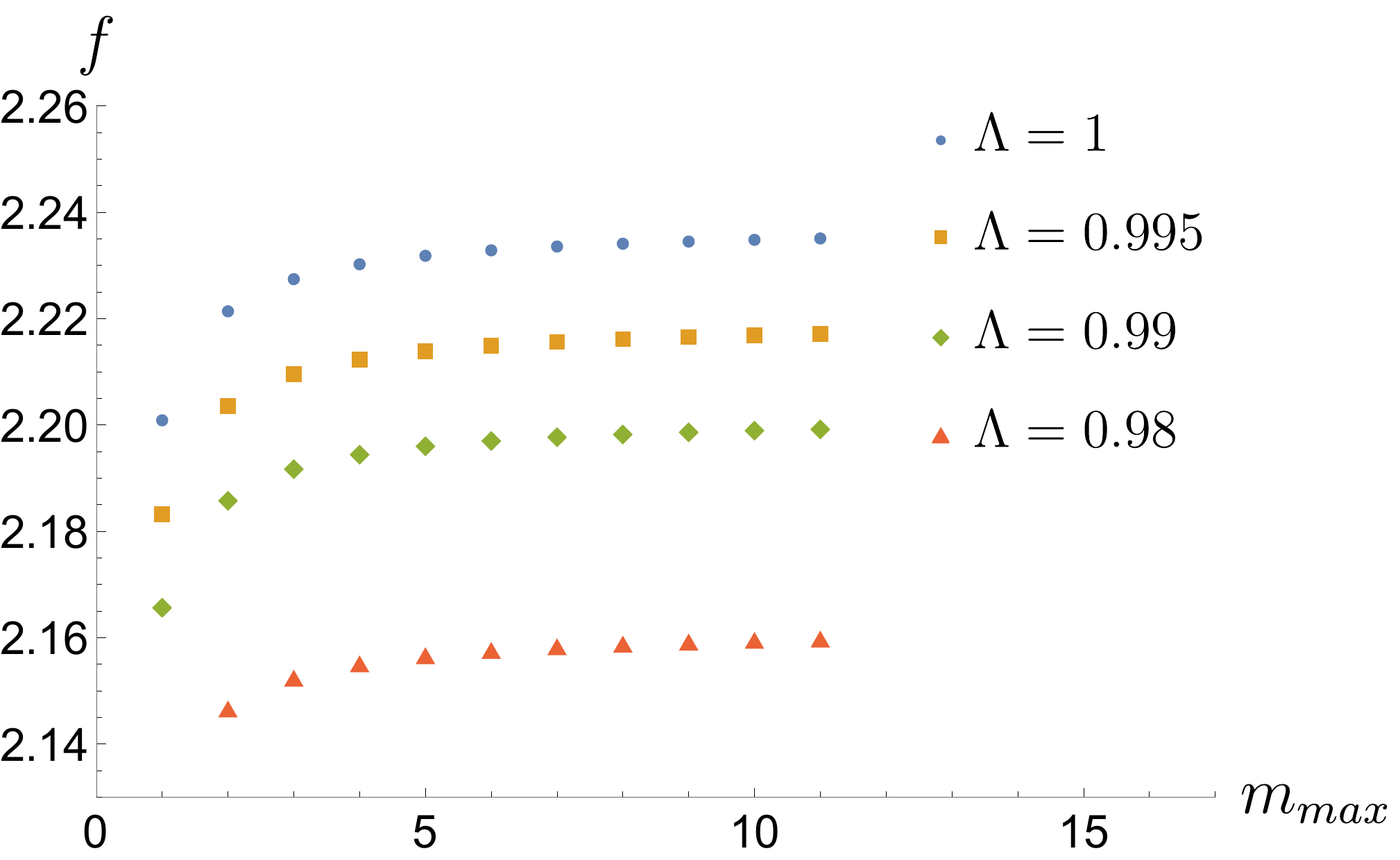}
\caption{Numerical evaluation of the function $f (\Lambda )$, with the sum truncated at a maximum value $m_{max}$ and different values of $\Lambda$. We observe a quick convergence even for small values of $m_{max}$.} \label{fregmax}
\end{figure}

\begin{figure}
\includegraphics[scale=0.45]{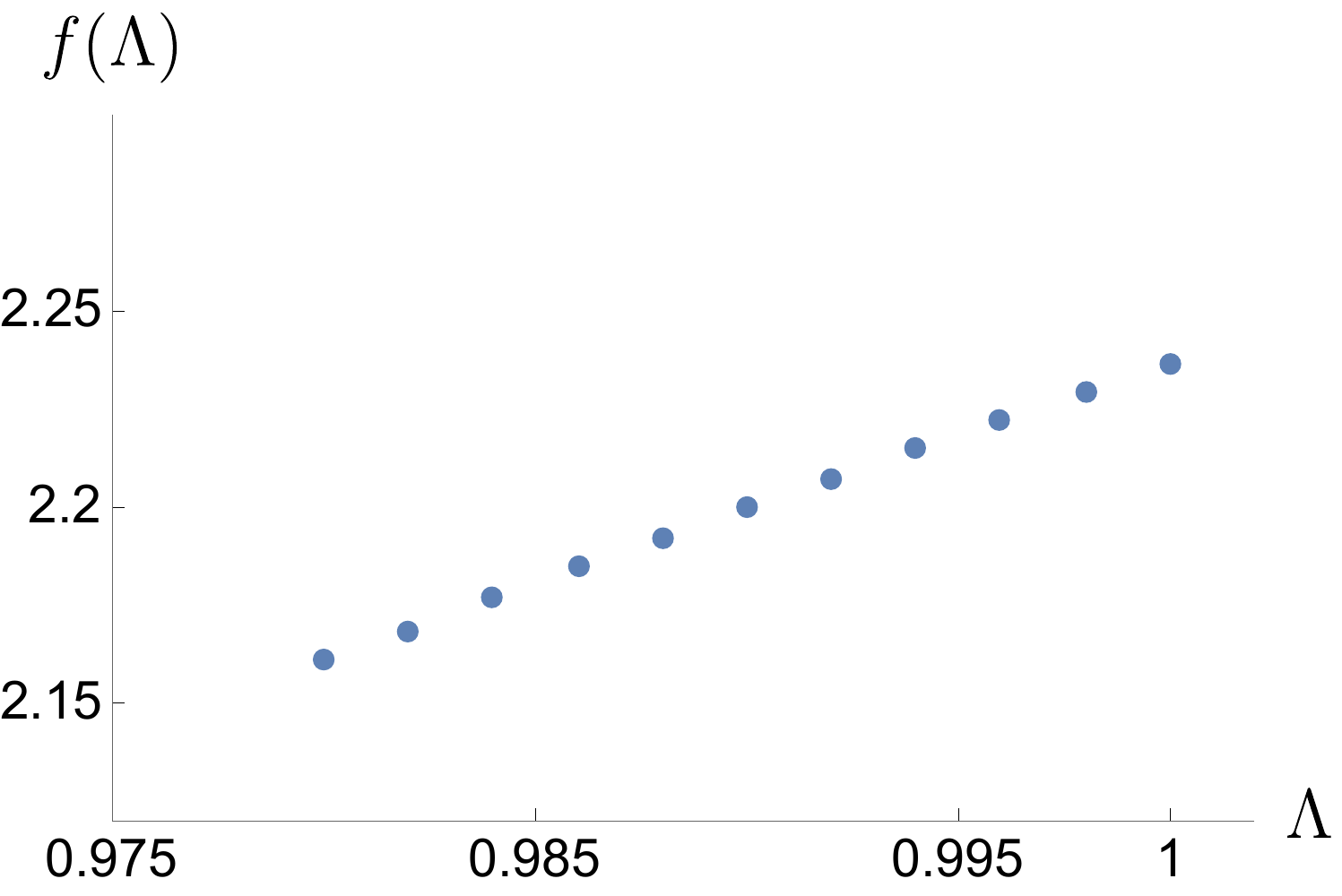}\hspace{0.5cm} \includegraphics[scale=0.39]{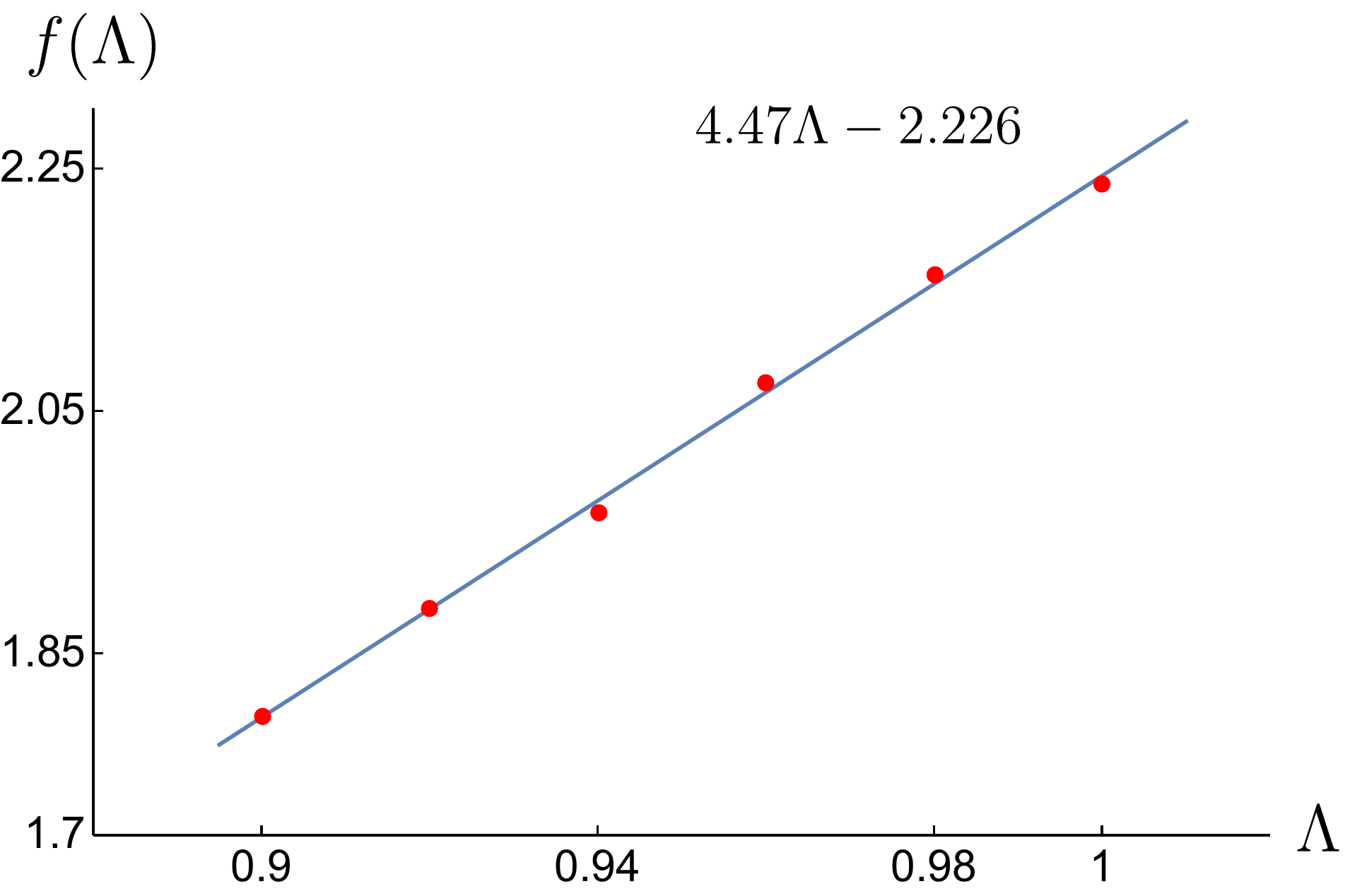}
\caption{The regularized Casimir pressure $f (\Lambda )$ in Eq. (\ref{f-regularize3}). In the region $\Lambda \in [0.9,\,1]$, $f (\Lambda )$ can be approximated by the linear function \mbox{$f (\Lambda )=4.47\Lambda  - 2.226$}.} \label{flambn}
\end{figure}

In practice, we compute the infinite sum occurring in (\ref{f-regularize3}) up to a maximum value $m=m_{max}$ without losing significant accuracy. The fast convergence of (\ref{f-regularize3}) is clearly seen in Fig. \ref{fregmax} where the function $f (\Lambda \approx 1)$ is displayed for $m_{max} \in [1,\,14]$.
Now, in Fig. \ref{flambn} the regularized Casimir pressure $f (\Lambda)$ with $m_{max}=20$ is presented for the range $\Lambda \in [0.9,\,1]$. We can see that it is a monotonically increasing function of the Lorentz-violating coefficient $\Lambda$. 
Making a simple linear interpolation of the data we obtain
\begin{equation}
f (\Lambda)  \approx  4.47 \Lambda  -  2.226 ,
\end{equation}
$0.9 \leq \Lambda \leq 1$. In the Lorentz invariant case $\Lambda=1$ with $m_{max}=8$, $m_{max}=40$ and $m_{max}=60$ we obtain $f(1)=2.2345$, $f(1)=2.2369$ and $f(1)=2.2371$, respectively.
As for the Lorentz-violating case $\Lambda=0.9$ with $m_{max}=8$, $m_{max}=40$ and $m_{max}=60$ we obtain $f(0.9)=1.7796$, $f(0.9)=1.7815$ and $f(0.9)=1.7817$, respectively.

\begin{figure}
\includegraphics[scale=0.23]{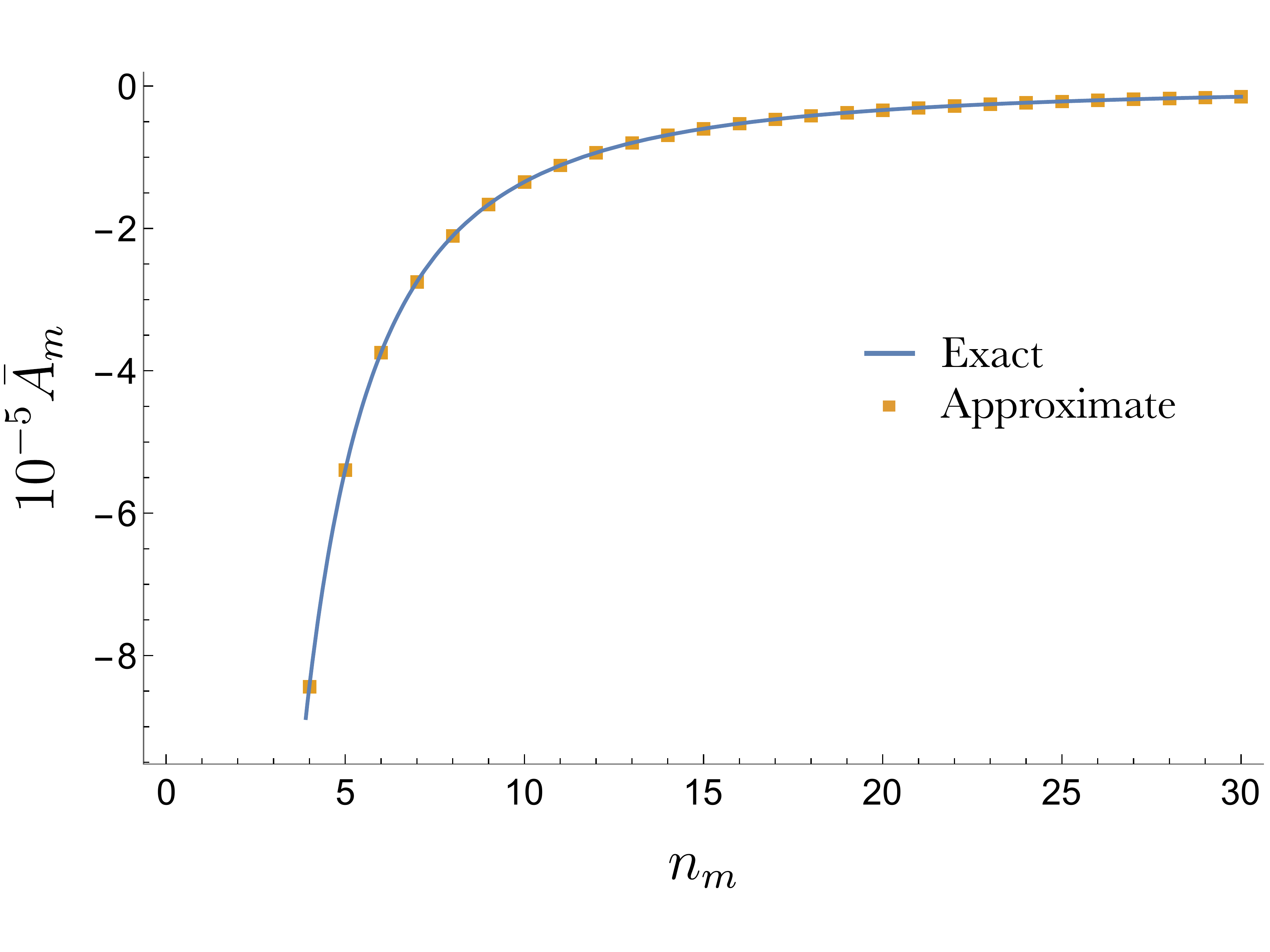}
\caption{Comparison between the exact expression (\ref{exact}) (continuous blue line) and the approximate expression (\ref{approximate}) (squared dots). We observe a perfect agreement even for small values of $n _{m}$.  } \label{Approximation}
\end{figure}

Up to a given precision, we can also approximate the expression for $f(\Lambda)$ in Eq. (\ref{f-regularize3}) as follows. By making use of the uniform asymptotic expansion of the modified Bessel functions (\ref{AsymptoticBessels}), we find that the product $I _{\mu} (\mu z) K _{\mu} (\mu z)$ behaves as \cite{Abramowitz}
\begin{align}
I _{\mu} (\mu z) K _{\mu} (\mu z) \sim \frac{t}{2 \mu } \left[ 1 + \frac{ 5 t ^{6} - 6 t ^{4} + t ^{2} }{ 8 \mu ^{2} } + \frac{ \left(385 t ^{4} - 462 t ^{2} + 81 \right) ^{2} t ^{4} }{1327104 \mu ^{4}} \right] , 
\end{align}
where $t = (1 + z ^{2}) ^{-1/2}$. Therefore, we deduce that, for $n _{m} \gg 1$, the coefficient (\ref{CoeffAm}) becomes
\begin{align}
\overline{\mathcal{A}} _{m} ( \Lambda ) &= - \Lambda \int _{0} ^{\infty} d x  \, \frac{ 2 n _{m} ^{2} x ^{3} \left( 81 x ^{4} - 300 x ^{2} + 4 \right) }{ \left( x ^{2} + 1 \right) \left[ 8 n _{m} ^{2} \left( x ^{2} + 1 \right) ^{3} + x ^{2} \left( x ^{2} - 4 \right) \right] }  \notag \\ & \hspace{3cm} \times \frac{ 48 n _{m} ^{2}  \left( x ^{2} + 1 \right) ^{3} \left( 27 x ^{4} - 254 x ^{2} + 104 \right) + 81 x ^{8} - 1062 x ^{6} + 3392 x ^{4} - 1256 x ^{2} - 16  }{ 1327104 n _{m} ^{4} \left( x ^{2} + 1 \right) ^{6} + 165888 n _{m} ^{2} x ^{2} \left( x ^{2} - 4 \right) \left( x ^{2} + 1 \right) ^{3} + \left(81 x ^{4} - 300 x ^{2} + 4 \right) ^{2}  } ,  \label{exact}
\end{align}
where we have changed the variable to $x = \Delta / m$. Now, we can power expand this expression to obtain
\begin{align}
\overline{\mathcal{A}} _{m} ( \Lambda ) &= - \Lambda \left( \frac{c _{1}}{n _{m} ^{2}} +  \frac{c _{2}}{n _{m} ^{4}} +  \frac{c _{3}}{n _{m} ^{6}} \right) \label{approximate}
\end{align}
where, up to order $10 ^{-6}$,
\begin{align}
c _{1} &=  \int _{0} ^{\infty} \frac{x^3 \left(27 x^4-254 x^2+104\right) \left(81 x^4-300 x^2+4\right)}{110592 \left(x^2+1\right)^7} dx = 0.00134654 , \\   c _{2} &= - \int _{0} ^{\infty} \frac{x^3 \left(81 x^4-300 x^2+4\right) \left(243 x^8-3282 x^6+10048 x^4-3736 x^2+16\right)}{5308416 \left(x^2+1\right)^{10}} dx = 0.0000525353 , \\   c _{3} &= \int _{0} ^{\infty} \frac{x^3 (81 x^4-300 x^2+4 ) (942597 x^{12}-16679034 x^{10}+91468224 x^8-173963088 x^6+58861392 x^4-188704 x^2-1664 )}{146767085568 (x^2+1 )^{13}} dx \notag \\ & = 2.62493 \times 10 ^{-6} .
\end{align}
In order to verify the validity of our approximate expression (\ref{approximate}) for the coefficients $\overline{\mathcal{A}} _{m} $, in figure \ref{Approximation} we plot the exact result (\ref{exact}) and the approximate result (\ref{approximate}), both scaled by a factor of $\Lambda ^{-1} \times 10 ^{5}$, for different values of $n _{m}$. This plot evinces that the approximate expression is valid up to $10 ^{-6}$. 

By numerical evaluation of Eq. (\ref{CoeffAm}) we can see that the leading order terms into the serie (\ref{f-regularize3}) corresponds to small values of $n _{m}$, as expected. To estimate their values, for the moment we take $\Lambda = 1$ and perform the numerical integrations. The results are presented in the Table \ref{table0}. This result  shows that if we want to evaluate the pressure to a precision of $10 ^{-4}$, for example, we can evaluate (\ref{f-regularize3}) by substituting the value of $\overline{\mathcal{A}} _{0}$, integrating numerically $\overline{\mathcal{A}} _{m}$ for $m=1,2,3,4 $, and asymptotically using (\ref{approximate}) for $m \geq 5$. The result is:
\begin{align}
f (\Lambda ) &= \overline{\mathcal{A}} _{0} (\Lambda ) + 2  \sum _{m = 1} ^{4}  \overline{\mathcal{A}} _{m} (\Lambda )  - 2 \Lambda  \sum _{m = 5} ^{\infty}  \left( \frac{c _{1}}{n _{m} ^{2}} +  \frac{c _{2}}{n _{m} ^{4}} \right) + \left( 2 \Lambda - 1 \right)  \ln (2 \pi ) , \label{f-regularize4}
\end{align}
It is easy to see  that in the Lorentz invariant case:
\begin{align}
f( 1 ) = 0.36331 + 2 (0.01676) - 2 \left[ (0.001346)(0.22132)  + (0.00005) (0.00357) \right] + 1.83 = 2.23413 , \label{fL1}
\end{align}
which is corroborated in Fig. \ref{fregmax} obtained by numerical evaluation of the function (\ref{f-regularize3}).

\begin{table}
\begin{center}
\begin{tabular}{| c | c | c | c | c | c | } \hline
$\overline{A} _{0}$ & $\overline{A} _{1}$ & $\overline{A} _{2}$ & $\overline{A} _{3}$ & $\overline{A} _{4}$  & $\overline{A} _{5}$ \\ \phantom{0}  0.36331 \phantom{0} & \phantom{0} 0.01060 \phantom{0} & \phantom{0} 0.00332 \phantom{0} & \phantom{0} 0.00172 \phantom{0} &  \phantom{0} 0.00111 \phantom{0} &  \phantom{0} 0.00080 \phantom{0} \\ \hline
\end{tabular}
\caption{Numerical values for the coefficient $\overline{A} _{m}$ for the lowest values of $m$.}
\label{table0}
\end{center}
\end{table}

\section{Conclusions}
\label{Conclusection}

%The Casimir effect is one of the most remarkable consequences of the nonzero vacuum energy predicted by quantum field theory which has been confirmed by experiments. Since it has been verified with astonishing high precision, it has become a testing ground for the predictions of new fundamental physical theories. Accordingly, the Casimir effect has been studied in different scenarios (e.g. Lorentz-violating theories) and geometries. This is precisely what motivates the present study. 

In this paper we have presented an analysis of the Casimir effect for a perfectly conducting cylindrical shell within the context of a Lorentz-violating scalar field theory. The effective field theory consists of the usual Klein-Gordon Lagrangian supplemented by the LV term $\lambda(u ^{\mu} \partial _{\mu} \phi ) ^{2}$, where $\lambda$ and the constant background four-vector $u^\mu=(u^0,\vec{u} \,  )$ control the intensity and direction of the breakdown of Lorentz symmetry, respectively. To compute the Casimir pressure, we evaluate the vacuum expectation value of the normal-normal component (to the cylindrical boundary) of the stress-energy tensor, for which we employ the point-splitting technique, which allows us to express the pressure in terms of the corresponding Green's functions. Here, we concentrate in two particular cases: the time-like and the radial space-like cases. In the former case, defined by $u ^{\mu}=(1, \vec{0})$, the usual Casimir pressure is exactly the same as in the Lorentz-symmetric case up to a scaling factor of $\sqrt{1  - \lambda }$. The later case, defined by $u ^{\mu} =(0,\vec{e} _{\rho})$, leads to a nontrivial modification of the Green's functions as well as the Casimir pressure, as can be seen in the general expression Eq. (\ref{f-regularize3}). The procedure to obtain the expression (\ref{f-regularize3}) is not simple at all, since as usual in the Casimir physics, divergences appear. For the regularization procedure we employ the Riemann zeta function technique. We find that the divergence appear by two different mechanisms: the product of a divergent sum and a divergent integral and the product of a divergent sum and a convergent integral. Following the standard techniques to assign finite values to such expressions, we find that only the former is nonzero, with the value $( 2 \Lambda - 1 ) \ln ( 2 \pi )$. This result correctly reduces to $\ln ( 2 \pi )$ when $\Lambda \to 1$, which is the correct value for the Lorentz-symmetric theory \cite{Milton}. After overcoming all technical difficulties, we get Eq. (\ref{f-regularize3}) for the dimensionless coefficient $f(\Lambda )$, which is the ratio between the Casimir pressure $\mathcal{F} (\Lambda )$ and $\mathcal{F} _{0} = - \hbar c / (8 \pi ^{2} R ^{4})$. We evaluate Eq. (\ref{f-regularize3}) by two routes. On the one hand, both the integral and sum of Eq. (\ref{f-regularize3}) are evaluated numerically. The result is displayed in the figure \ref{fregmax} for different values of $\Lambda$. On the other hand, we observe that the main contributions to the sum in Eq. (\ref{f-regularize3}) come from lowest values of $m$. Therefore, we obtain an asymptotic expression for the coefficient $\overline{A} _{m}$ for large $m$, which reproduces the exact result up to a precision of $10 ^{-6}$. Also, by analyzing the order of magnitude of $\overline{A} _{m}$ for small $m$, we split the sum (\ref{f-regularize3}) into small values of $m$ (from 1 to 4 for a precision to $10 ^{-6}$), and larger values $m \geq 5$. The first part is computed numerically, while the later is obtained from the approximate expression (\ref{f-regularize3}). We confirm the validity of our approximate expression (\ref{f-regularize4}) by comparing the numerical result for $\Lambda = 1$, given by Eq. (\ref{fL1}), with those presented in the figure \ref{fregmax}.

Is worth mentioning that the Lorentz-violating scalar field theory under consideration was originally conceived within the scalar sector of the Standard-Model Extension. However, since no deviation from Lorentz symmetry has been detected in particle physics, it is expected that physical relevant systems correspond to  $\| \lambda u ^{\mu} u ^{\nu} \| \ll 1$, for any $\mu$ and $\nu$. Lorentz-violating models also appear in connection to Riemann-Finsler spacetimes \cite{Edwards} where the notion of distance is controlled by additional quantities beyond the Riemann metric, which can intuitively play the role of Lorentz-violating coefficients \cite{KOSTELECKY2011137}. The model in the present work is a particular case of those presented in Ref. \cite{Edwards} through the identification $(\hat{k} _{c} ) ^{\mu \nu} = \lambda u ^{\mu} u ^{\nu}$.  There are also scenarios in which Lorentz symmetry is naturally broken, such as condensed matter systems, where Lorentz-violating coefficients are not much smaller than one. For example, the intrinsic anisotropy in a crystalline solid can play the role of the constant background, and this manifest in the optical and electronic properties of the material. On this subject, it has been found that the Lorentz-violating coefficients in the fermion sector of the SME can be identified with different properties of the low-energy description of topological phases, such as the tilting of the cones and the chiral mass term. This provides an interesting platform in which Lorentz-violating fermionic theories can be tested with material media. In a similar fashion, it would be interesting to study if the optical properties of some material media can be described by the above discussed scalar field theory. In this regard, it is known that axion electrodynamics \cite{Wilczek} can be derived from the CPT-odd photon sector of the  minimal SME defined by the Lagrangian $\mathcal{L} _{AF} = \frac{1}{2} (k_{AF}) ^{\kappa} \epsilon _{\kappa \lambda \mu \nu} A ^{\lambda} F ^{\mu \nu} $, and it has found many applications in the description of the electromagnetic response of topological phases \cite{MCU1, MCU2, MCU3, MartinCyl, MCU5}. In particular, the Casimir effect between topological insulating systems has been considered in Refs. \cite{Cortijo1, Cortijo2, Grushin, MCU4}. Thus, Lorentz-violating electrodynamics is potentially testable within these new phases of matter. Along the same line, the scalar field theory defined by the Lagrangian (\ref{Lagrangian1}) can be derived, from the CPT-even photon sector of the  minimal SME defined by the Lagrangian $\mathcal{L} _{F} = \frac{1}{2} (k_{F})_{\kappa \lambda \mu \nu} F^{\kappa\lambda} F ^{\mu \nu} $. Up to our knowledge, there is no material media whose optical response is described by the Lagrangian $\mathcal{L} _{F}$, and hence the model may remain not testable within the condensed matter framework.

\acknowledgements{A.M.-R. has been partially supported by DGAPA-UNAM Project No. IA101320 and by Project CONACyT (M\'{e}xico) No. 428214. C. A. E. is partially supported by the Project PAPIIT No. IN109321. A.M.E-R. is supported in part by CONACyT grant 237351 (Mexico).}

\bibliography{references.bib}
\end{document}